\documentclass[twocolumn,a4paper,superscriptaddress,floatfix,showpacs,aps]{revtex4}%
\usepackage[english]{babel}
\usepackage[utf8]{inputenc}
\usepackage{amsmath}
\usepackage{graphicx,epstopdf}
\usepackage{amssymb,epsfig,color,textcase}
\usepackage{hyperref}
\usepackage{amsfonts}
\usepackage{amssymb}
\usepackage{ulem}
\usepackage{graphicx}%

\providecommand{\U}[1]{\protect\rule{.1in}{.1in}}

\begin{document}

\title{Orbital physics in transition metal compounds: new trends}

\author{Sergey V. Streltsov}
\email{streltsov@imp.uran.ru}
\affiliation{M.N. Miheev Institute of Metal Physics of Ural Branch of Russian Academy of
Sciences, 620137, Ekaterinburg, Russia}
\affiliation{Theoretical Physics and Applied Mathematics Department, Ural Federal
University, Mira St. 19, 620002 Ekaterinburg, Russia}

\author{Daniel I. Khomskii}
\affiliation{II. Physikalisches Institut, Universit$\ddot a$t zu K$\ddot o$ln,
Z$\ddot u$lpicher Stra$\ss$e 77, D-50937 K$\ddot o$ln, Germany}
\date{\today}

\begin{abstract}
In the present review different effects related to the orbital degrees of freedom are discussed. Leaving aside such aspects as the superexchange mechanism of the cooperative Jahn-Teller distortions and different properties of ``Kugel-Khomskii''-like models, we mostly concentrate on other phenomena, which are in the focus of modern condensed matter physics. After a general introduction, we start with the discussion of the concept of effective reduction of dimensionality due to orbital degrees of freedom and consider such phenomena as the orbitally-driven Peierls effect and the formation of small clusters of ions in the vicinity of the Mott transition, which behave like ``molecules'' embedded in a solid. The second large section is devoted to the orbital-selective effects such as the orbital-selective Mott transition and the suppression of magnetism due to the fact that part of the orbital start to form singlet molecular orbitals. At the end the rapidly growing field of the so-called ``spin-orbit-dominated'' transition metal compounds is briefly reviewed including such topics as the interplay between the spin-orbit coupling and Jahn-Teller effect, the formation of the spin-orbit driven Mott and Peierls states, the role of orbital degrees of freedom in generation of the Kitaev exchange coupling, and the singlet (excitonic) magnetism in 4d and 5d transition metal compounds.
\end{abstract}

\maketitle
\tableofcontents

\section{Introduction}
Systems with strongly correlated electrons, in particular transition metal (TM) compounds, present a very interesting class of materials with extremely rich properties, see e.g. \cite{khomskii2014transition,Imada1998}. There are among them metals, insulators (of a special kind), systems with metal-insulator transitions; they display different types of ordering (magnetic, charge ordering (CO), orbital ordering (OO)), cooperative Jahn-Teller effect, and last but not least, high-temperature superconductivity.  All this richness is mainly caused by strong electron correlations and by the presence in them, and mutual interplay of different degrees of freedom: charge, spin, orbital, and all this of course on the background of the lattice, with  which all these electronic degrees of freedom often strongly interact.

The crucial general feature of these systems is a fundamental importance of electron-electron interaction, which determines the main properties  of these systems, changes behaviour of electrons as compared with the standard free-electron-like or band description, and leads to localization of electrons on respective sites. These are the famous Mott, or Mott-Hubbard insulators. Most often such electron localization leads to the appearance  of the localized magnetic moments, which then determine all the rich magnetic properties of Mott insulators, and sometimes also gives charge ordering. The existence of two different limiting cases – strongly correlated and localized electrons, vs weakly-interacting itinerant ones, leads also to the possibility of a transition between these states, caused by change of temperature, pressure, doping etc. – the famous Mott metal-insulator transitions (MIT).

Besides charge and spin degrees of freedom, in real TM compounds one has  to take into account  also orbital degrees of freedom, which lead to many nontrivial properties – orbital ordering, directional character of many properties, nontrivial effects related to the relativistic spin-orbit coupling (SOC). All these effects taken together can lead to novel, very interesting phenomena, which are the subject of the present review. For example, the directional character of orbitals may result in spontaneous reduction of the dimensionality, when three-dimensional system like KCuF$_3$ or Tl$_2$Ru$_2$O$_7$ starts to behave like  one-dimensional magnets. This effect together with another very interesting phenomenon - formation of small clusters, where electrons are practically delocalized, while a system as a whole is still insulating, are discussed in Sec.~\ref{sec:red-dim}. This concept of ``molecules in solids'', leading to a ``step-wise'' Mott transition is an alternative to a homogeneous Mott transition. Another important aspect of the Mott physics is discussed in Sec.~\ref{sec:OS-large}. This is the so-called orbital-selective Mott transition, when due to direction character of the orbitals there is a substantial overlap between some of orbitals centered on different sites, while hopping (and hence the bandwidth) between others is much smaller so that they turn out to be more susceptible to the Mott transition, which again occurs stepwise, but in momentum space, first for narrow and then for wider bands. Moreover, even if there is no Mott transition, this separation on more and less ``localized''  orbitals (in fact these are the electrons, not orbitals, which can be localized or itinerant) may strongly affect magnetic properties of a system resulting, in particular, in a suppression of the double exchange mechanism of  ferromagnetism. Finally there is a large group of effects related to the spin-orbit coupling (see Sec.~\ref{sec:SOC}), which is under detailed study right now and which has already brought up such phenomena as the spin-orbit assisted Mott state, the Kitaev and excitonic magnetism.

The present review is devoted to  a general description of the main concepts of orbital physics, with the central  attention paid to the new development in this big field. We describe novel phenomena mentioned above and also discuss many real examples of systems the properties of which find natural explanation using these concepts. For completeness, to make our review more self-contained, we also included in the first two introductory sections a general description of the main concepts in the field of systems with strongly correlated electrons, in particular TM compounds, paying main attention to the role of orbital degrees of freedom in different phenomena. More complete presentation of this material one can find in many monographs and textbooks, in particular in \cite{Goodenough,fazekas,Khomskii-book1,khomskii2014transition}, and in review articles\cite{Khomskii1970,KK-UFN,Zaitsev1986,Izyumov1995} . We also do not discuss here in details possible types and mechanisms of orbital ordering and extensive literature devoted to ``Kugel-Khomskii'' and compass models. The first topic is reviewed in the rather old but still not obsolete paper\cite{KK-UFN} and in the recent book\cite{khomskii2014transition}. For other aspects of the orbital physics we may recommend reviews \cite{Nussinov2015,Oles2012}.

\section{Basic concepts of describing electrons in solids}
\label{sec:basic-concept}
To start with, we discuss at the beginning  general ways to describe the state of electrons in solids in different situations. The simplest approach, from which the description of electrons is always started, is that of free electrons (band structure theory). In this type of treatments one considers the motion of an electron in a periodic lattice potential, first ignoring electron-electron interaction or treating it in a mean-field way. This leads to the well-known formation of energy bands – the regions of  allowed states, in general divided  by the forbidden regions – energy gaps.

There exist two main approaches for description of these energy bands: the weak coupling approximation, in which  periodic potential of the lattice is treated as a perturbation, and the tight-binding approximation. For our purposes, in particular for describing $d$-electrons of TM compounds, the second method is more useful, and we will mostly use it below. When we start with the band description, we can easily get both insulating and metallic states by filling available band states. According to Pauli principle we put two electrons with spins up and  down at each state. If some bands turned out to be partially filled, we will then  deal with a metal, like Na or Al. And if some bands will be completely filled, and the upper-lying bands separated from the occupied ones by an energy gap are empty, we will have band insulator or semiconductor like Ge or Si.

\begin{figure}[tbp]
   \centering
  \includegraphics[width=0.49\textwidth]{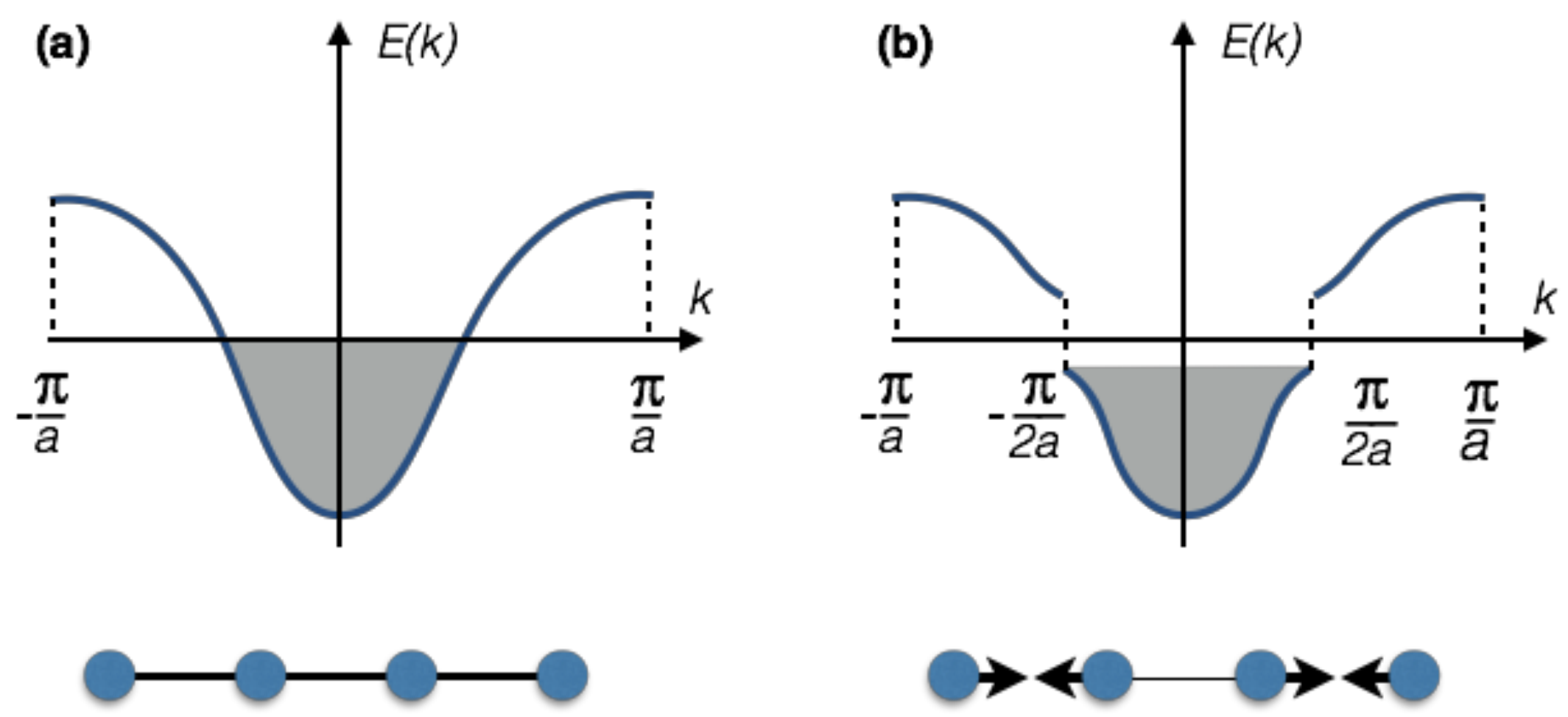}
  \caption{\label{Peierls}  Peierls transition accompanied by opening of the gap in the electronic spectrum $\varepsilon (\vec k)$. Distance between sites in the uniform chain is $a$.}
\end{figure}

In the tight-binding approximation we can speak of bands which are formed by intersite hopping of electrons between particular ionic states, e.g. $1s$ states of hydrogen or $3d$ states of TM ions. For the lattice of $N$ sites each such (nondegenerate) band would contain $N$ electronic states, into each of which we can put two electrons, so that there are places for $2N$ electrons in such band, e.g. $1s$ band of a lattice made of equally-separated hydrogen ions (protons), see Fig.~\ref{Peierls}a. Corresponding tight binding electrons can be described by the Hamiltonian
\begin{eqnarray}
\label{kin-energy-R}
H = -t \sum_{\langle ij \rangle \sigma} c^{\dagger} _{i\sigma} c_{j\sigma},                                                                              
\end{eqnarray}
where $c^{\dagger} _{i\sigma}$ and $c_{j\sigma}$ are creation and annihilation electron operators on sites $i$ and $j$, $\sigma$ is the electron spin. The intersite hopping matrix element $t$ is positive for $s$ orbitals. Summation in \eqref{kin-energy-R} goes over all inequivalent pairs of nearest neighbor lattice sites, numerated by indexes $i$ and $j$. In momentum space the Hamiltonian reads as 
\begin{eqnarray}
\label{kin-energy-k}
H =  \sum_{ {\bf k} \sigma} \varepsilon ({\bf k}) c^{\dagger} _{{\bf k} \sigma} c_{{\bf k} \sigma}, 
\end{eqnarray}
with the  dispersion $\varepsilon ({\bf k}) = -2t \cos (k_x a)$. If we have noninteger or odd number of electron per site, e.g. one electron as for hydrogen lattice, the band will necessarily be partially- (e.g. half-) filled, and we will have a metal. Only if we have even number of electrons per site, such system will be a usual band insulator (and even in this case in realistic situation we can have a metal or semimetal, if some bands overlap).

In this band picture one can also have metal-insulator transitions. These may be caused by structural transitions with the  lattice distortions, which  opens a gap exactly at the Fermi-surface. The simplest example of that is the Peierls transition in a one-dimensional (1D) case. If we have for example a regular chain of sites (e.g. hydrogen atoms) with, say, one electron per site, then $1s$ band would be half-filled, see Fig.~\ref{Peierls}a. Dimerization of this chain (a first step towards the formation of $H_2$-molecules from the lattice of hydrogens) would open the gap exactly at the Fermi-surface and would decrease electron energy, see Fig.~\ref{Peierls}b, and this decrease overcomes the loss of lattice (deformation) energy, i.e. such chain would be always unstable to the dimerization (see also review \cite{Bulaevskii1975}). We note right away that such instability in a chain would exist not only for half-filled band, but also for other fillings: e.g. for the band filled by 1/3 or 2/3 we would have trimerization, and for the 1/4 filled band – tetramerization. We will see real example of such phenomena later on, in Sec.~\ref{sec:orb-Peierls}.

Such metal-insulator transitions in the band picture may occur not only for (quasi)-1D systems, but also in a more general situation. The usual condition for that is the so-called nesting of the Fermi-surface,  at which some parts of the Fermi-surface coincide when shifted by a certain vector ${\bf Q}$. In this case the superstructure with this wave vector ${\bf Q}$ could be formed – charge density wave (CDW) in case of effective electron attraction (e.g. via phonons) or spin density wave (SDW) for electron repulsion. And if the gap which opens at these transitions would cut the whole Fermi-surface, it would lead to a metal-insulator transition. Such examples are met in some TM dichalcogenides such as, e.g.  TaS$_2$.

Now, according to the band picture, if there is no dimerization, the regular lattice of hydrogen atoms with one electron per site should have half-filled band and should be metallic, irrespective of the distance between atoms, or of the value of the intersite hopping matrix element $t$, which for large distance between sites would be exponentially small. Of course, this is beyond common sense: we should rather deal here with a collection of neutral (hydrogen) atoms with electrons localized one per site.

The reason for this was explained already long ago\cite{Mott1937}, see also Appendix A.1 in Ref.~\cite{khomskii2014transition}: when we remember that electrons repel each other, it immediately becomes clear that if we start with one electron per site and then try to create charge carriers, transferring an electron from this site to the other one, the repulsion of the transferred electron with the ``its own'' one, already existing at this site, will prevent such charge transfer. In effect the material would become an insulator with electrons localized each at its own site. This is what we now call Mott, or Mott-Hubbard insulators. And, in contrast to the band insulators, described at the beginning if this section, the very fact that such system remains insulating is due to electron-electron interaction, and not due to the interaction of independent electrons with the periodic lattice potential.

To treat this state we have to generalize the description presented in Eqs. \eqref{kin-energy-R}-\eqref{kin-energy-k} and include electron-election interaction – at least the Coulomb repulsion at the same site. Corresponding model
\begin{eqnarray}
\label{Hubbard-model}
H = -t \sum_{\langle ij \rangle \sigma} c^{\dagger} _{i\sigma} c_{j\sigma} + U \sum_i  n_{i\uparrow} n_{i\downarrow},                                                                              
\end{eqnarray}
where $n_{i \sigma} = c^{\dagger}_{i \sigma} c_{i \sigma}$ is the electron density, is called the Hubbard model, and it serves nowadays as the basic model to describe the physics of systems with strong electron-electron interaction, or with strong electron correlations.

According to the physics discussed above and described by the Hubbard model \eqref{Hubbard-model}, the state of the system is characterized by two parameters: the average electron density $n = N_{el}/N$  and the effective interaction $U/t$, or $U/W$, where $W = 2zt$ is the electron bandwidths (for simple lattices like linear chain, square or cubic lattice; $z$ is the number of nearest neighbors). Here $N$ is the number of sites and $N_{el}$ is the number of electrons. If $U/t \ll 1$, we are dealing with weakly interacting electrons, and in this case the standard band description is valid; electron-electron interaction can then be  taken into account by perturbation theory, using for example Feynman diagram technique, etc. Also for $n \ne 1$  we would have a metal – although for strong interaction $U \gg t$ it could be a special type of ba metal, with still strong correlations (such metallic state could be in principle rather fragile and very sensitive and unstable to any extra perturbations – longer range interactions, etc.). However at least in simplest cases  this description catches the main physical effect: the creation of a novel state – Mott insulator with localized electrons  for half-filled bands (one electron per site $n=1$) and for strong interaction $U/t \gg1$. And we see that in this state we simultaneously create localized magnetic moments: each electron localized at a respective site gives a localized spin moment, corresponding to $S=1/2$.

When in this situation we take into account only the dominant term in the Hamiltonian \eqref{Hubbard-model}, the interaction term $Un_{i\uparrow} n_{i\downarrow}$, the spin direction would not matter, and the system would be paramagnetic (with disordered localized spins). However, if we also consider electron hopping, the first term in \eqref{Hubbard-model}, this hopping lifts spin degeneracy in the second order of perturbation theory in $t/U \ll 1$, and it leads to the antiferromagnetic interaction of localized spins $\sim t^2/U$, i.e. the low-energy states of the system can be effectively described by the Heisenberg model (see Sec.~\ref{sec:GKA} for details)
\begin{eqnarray}
\label{eq:Heisenberg}
H = J \sum_{ij} \mathbf {S}_i \mathbf S_j    = \frac{2t^2}{U} \sum_{ij} \mathbf S_i \mathbf S_j,                                                                                                                                                                                                                                                     
\end{eqnarray}
where ${\bf S}_i$ is the spin operator acting at a site $i$, and $J$ is the exchange coupling between spins on two  such sites (and hence in principle it can be different for different pairs, i.e. $J \to J_{ij}$ in this situation). The ground state of such system would be a Mott insulator with antiferromagnetic spin ordering. For only two sites and two elctrons we would then have the singlet ground state
\begin{eqnarray}
\label{eq:HL}
\Psi_{HL} = \frac1{\sqrt{2}} \left(c^{\dagger}_{1\uparrow} c^{\dagger}_{2\downarrow} - c^{\dagger}_{1\downarrow} c^{\dagger}_{2\uparrow} \right) | 0 \rangle.                                                                    
\end{eqnarray}
This is what is called in the theory of chemical bond the Heitler-London (HL) description. 

Here we should say that actually also for noninteracting electrons described by the simple Hamiltonian \eqref{kin-energy-R} the ground state would be also a unique singlet state – a filled Fermi-surface, in which there are two electrons with spins up and down at every occupied state. For only two such sites the ground state would also be a singlet
\begin{eqnarray}
\label{eq:MO}
\Psi_{MO} = \frac12
 \left(c^{\dagger}_{1\uparrow} + c^{\dagger}_{2\uparrow} \right)
 \left(c^{\dagger}_{1\downarrow} + c^{\dagger}_{2\downarrow} \right) | 0 \rangle
\end{eqnarray}                                                                      
Such state in the theory of chemical bonds is called the Hund-Mulliken, or molecular orbital (MO) state (sometimes denoted as MO LCAO: Molecular Orbitals – Linear Combination of Atomic Orbitals). 

In quantum chemistry it was relatively soon realized that both MO \eqref{eq:MO} and HL \eqref{eq:HL} wave functions describe just two limiting cases, and for realistic calculations one should rather use a linear combination of homopolar states, given by the HL wave function \eqref{eq:HL}, and ionic contributions $c^{\dagger}_{1\downarrow} c^{\dagger}_{1\uparrow} + c^{\dagger}_{2\downarrow} c^{\dagger}_{2\uparrow}$, but (in contrast to MO) with a variational coefficient:
\begin{eqnarray}
\label{eq:Coulson}
\Psi_{CF} &=& 
\frac{\sin \theta}{\sqrt 2} \left(c^{\dagger}_{1\uparrow} c^{\dagger}_{2\downarrow} - c^{\dagger}_{1\downarrow} c^{\dagger}_{2\uparrow} \right) | 0 \rangle \nonumber \\
 &+& \frac{\cos \theta}{\sqrt 2} \left(c^{\dagger}_{1\downarrow} c^{\dagger}_{1\uparrow} + c^{\dagger}_{2\downarrow} c^{\dagger}_{2\uparrow} \right) | 0 \rangle.
\end{eqnarray}
The wave function in the form $\Psi_{CF}$ is often called Coulson-Fisher wave function\cite{Coulson1949}. These notions will be very important for our discussion in the main body of this paper.

We thus see that the dichotomy  between  two descriptions of chemical bonds in molecules – MO and HL – have exact counterpart in two types of solids: those with itinerant electrons described by the band theory, and localized electrons in Mott insulators due to strong electron correlations. But, in contrast to the case of molecules, where with the increase of electron correlations we continuously go over from the MO to HL description, cf. for example  the Coulson-Fisher form \eqref{eq:Coulson}, for large concentrated solid these two states are really two different thermodynamic states of matter, with sharp, well defined transition between them – the Mott transition. This transition can be caused simply by the change of a parameter $U/t$ (which can be in many systems reached experimentally under pressure, which leads to the increase of electron hopping $t$, but in some very interesting cases also by change of temperature, doping etc). And the properties of a system close to this localized-itinerant crossover turned out to be very interesting and nontrivial, with some rather unexpected features emerging, see Sec.~\ref{sec:OS}.

\section{Basic effects related to the orbital degrees of freedom\label{sec:RealDesc}}

\subsection{Crystal-field splitting, spin-state transition \label{sec:CFS}}
\begin{figure}[t]
   \centering
  \includegraphics[width=0.49\textwidth]{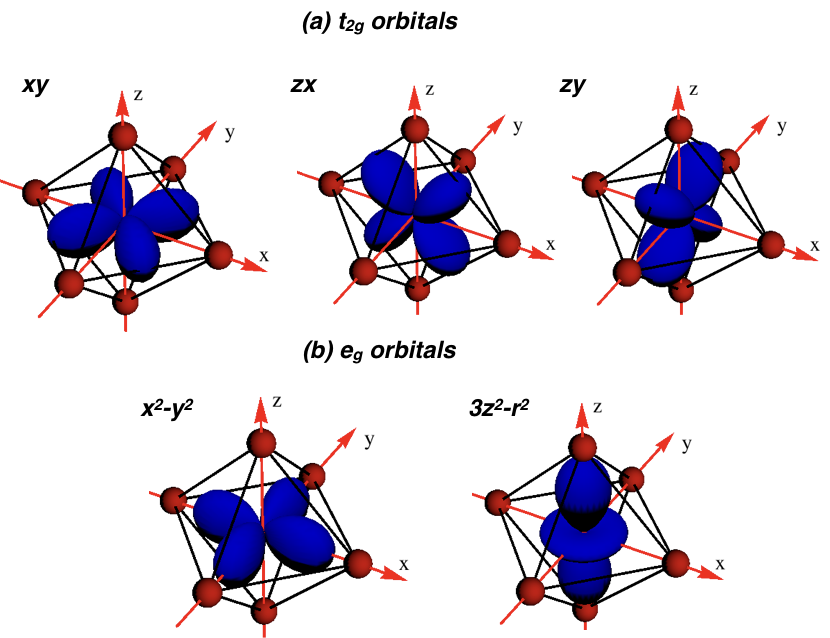}
  \caption{\label{fig:cubic-orbitals} Cubic harmonics corresponding to $d$-orbitals  in octahedral surrounding.}
\end{figure}

When we want to apply these general ideas to TM compounds, several important ingredients have to be included, which make the description, on one hand, more realistic, and which, on the other hand, often lead to novel phenomena. The most interesting (from the point of view of physical properties) TM compounds have partially-filled $d$-bands. Five $d$-states are degenerate in isolated atoms or ions, but become split when put the ion in a crystal. In fact one has to classify these split states according to corresponding irreducible representations. Thus if a TM ion turns out to be inside ligand octahedra, as it often happens in TM compounds (e.g. NiO, La$_2$CuO$_4$, LaCoO$_3$ etc.), its $d$-levels are split into the $t_{2g}$ and $e_g$ sub-shells, see Fig.~\ref{jj-d4-d5} below: the $e_g$ orbitals are directed as much as possible towards the ligands, while the $t_{2g}$ ``look'' in between of them, see Fig. \ref{fig:cubic-orbitals}.  This effect is called crystal-field splitting.

There are two equally important contributions to the crystal-field splitting. First of all, there is indeed an effect of the electric field created by a local surrounding. Negatively charged ligands repel negative electron charge density corresponding to $d$-orbitals. This repulsion is larger for the $e_g$ orbitals directed to the ligands, and these orbitals go higher in energy than the $t_{2g}$ ones. However, there is also another contribution to the crystal-field splitting, that due to a hybridization with the ligand $p$ orbitals. In conventional TM compounds the ligand $p$ orbitals lie lower in energy than the $d$-orbitals of TMs, and the hybridization between them shifts $d$-orbitals even higher. In the case of octahedral surrounding this shift will be larger for the $e_g$ orbitals. So we see that both effects usually work hand in hand. 

However, there can be exceptions from this rule, e.g. if at least some of ligand $p$ states lie {\it higher} than $d$, then the hybridization will push these $d$-states down, while Coulomb forces push them up (both stronger for $e_g$ orbitals). If hybridization wins the bonding states of $e_g$ symmetry (these will be a mixture of $p$ and $d$-orbitals) will be {\it lower} than those of $t_{2g}$. This happens in Cs$_2$Au$_2$Cl$_6$\cite{Ushakov2011}, and it can be also  expected in other  systems with a negative charge-transfer gap, $\Delta_{CT}$, which is the energy cost for the reaction $d^n p^6 \to d^{n+1}p^5$\footnote{If $\Delta_{CT}>0$, as in normal TM oxides, then we loose the energy transferring electrons from a ligand to a metal, while if $\Delta_{CT}<0$, such  transfer occurs spontaneously to minimize the total energy of a system.}.  A negative charge-transfer regime can be realized in the case of unusually high oxidation state of a TM, when a system cannot afford such a strong charge redistribution between a metal and a ligand as the chemical formula requires. This is why there are holes in the ligand $p$ orbitals, which  appear to be higher in energy than some of the $d$-orbitals in these systems. Thus for example in CrO$_2$, where Cr is nominally 4+ and O is 2-, it is rather unfavorable to transfer four electrons from Cr ion to O; instead there appear holes in the O $2p$ band, with Cr being practically 3+ \cite{Korotin1998}. 

The crystal-field splitting ($\Delta_{CFS}$) often has a dramatic influence on the magnetic properties of TM compounds. From atomic physics we know that the Hund's rules determine the state with partially-filled levels. Simply stated, they tell that the state of a many-electron system should be such that, first, the total spin $S_{tot} = \sum_i s_i$, and then the total orbital moment $L_{tot} = \sum_i l_i$ of an ion should be maximum possible. This in particular means that, e.g., Co$^{3+}$ ion with configuration $3d^6$ should have $S_{tot}=2$. However, this is not always the case. When a TM ion is put in an octahedral surrounding, the ligand crystal-field splits its $d$-shell, making filling of higher lying $d$-levels (the $e_g$ levels in octahedra) energetically unfavorable, which may result in violation of the first Hund's rule. A classical example of such a situation is LaCoO$_3$, where the spin-state of Co$^{3+}$ is the low-spin $S_{tot}=0$  (electron occupation $t_{2g}^6$). and the transition (known in chemistry as spin crossover) from  the low-spin  ($S_{tot}=0$, $t_{2g}^6 e_g^0$) to the intermediate-spin ($S_{tot}=1$,  $t_{2g}^5 e_g^1$) or to a  mixture of low-spin and high-spin ($t_{2g}^4 e_g^2$) states occurs\cite{GoodenoughLaCoO3,KorotinLaCoO3,Haverkort2006,Kunesh2011}. 

It is often sufficient describe the Hund’s rule in the mean-field approximation by the following Hamiltonian: 
\begin{eqnarray}
\label{eq:Hund}
H_{Hund} = - J_H \sum_{m \ne m'} \left( \frac 12 + 2 S_m^z S_{m'}^z \right),
\end{eqnarray}
where $m,m'$ numerate orbitals, and  $J_H$ is the intra-atomic Hund’s exchange parameter. It is easy to see that if one uses this  Hamiltonian, then in order to find the Hund’s exchange energy for each atomic configuration one needs simply to count the number of nonequivalent pairs of electrons with parallel spins  (e.g., for Co$^{3+}$ the low-spin state will have $E_{Hund} = - 6 J_H$, intermediate spin $E_{Hund} = - 7 J_H$, and high-spin state $E_{Hund} = - 10 J_H$). 

Spin state transitions can be found in many other TM compounds based, in addition to Co$^{3+}$, on Fe$^{2+}$, and more rarely Fe$^{3+}$, Mn$^{2+}$, and Mn$^{3+}$ ions. It is rather important to mention two points in this regard.  First of all, the spin-state transitions are more typical for the $3d$ and not for $4d$ and $5d$ TM compounds. In $3d$ systems the $t_{2g}-e_g$ splitting is $\Delta_{CFS}\sim$1.5-2 eV and it can easily compete with the intra-atomic exchange interaction, which is given by $J_H \sim 1$ eV and which arranges electrons according to the Hund's rule. In contrast, due to larger principal quantum number, the  $4d$ and $5d$ orbitals are more spatially extended than $3d$\cite{Goodenough}. As a result both kinetic and Coulomb contributions to the crystal-field splitting are larger and the $t_{2g}-e_g$ splitting exceeds 3-4 eV in the systems based on these ions\cite{Streltsov2012a}. 
In effect $4d$ and $5d$ elements typically assume low-spin states, putting as many electrons as possible into the lower-lying $t_{2g}$ levels.

While one cannot completely rule out the possibility that even in this case there may occur  spin-state transitions within $t_{2g}$ levels split by noncubic crystal field,  generally this is rather unlikely since corresponding splitting is typically much smaller than $J_H$. Indeed, the attempts to describe the properties of some materials by the spin-state transition due crystal-field splitting of the $t_{2g}$ subshell, e.g. \cite{Khalifah2002}, failed \cite{Wu2006,Zhou2012,Streltsov2013a}. One might expect, though, that this idea may apply to some early $5d$ TM compounds, where $J_H$ is expected to be as small as 0.3-0.5 eV, and $\Delta_{CFS}$ within the $t_{2g}$ subshell due to noncubic crystal field  can be also $\sim$0.5 eV

\subsection{Orbital degrees of freedom and magnetism\label{sec:GKA}}
While the crystal-field splitting in the $t_{2g}$ or $e_g$ subshells (not the main splitting between $t_{2g}$ and $e_g$!) is unlikely to lead to a spin-state transition, it (and all the more the ``main’’ $t_{2g}-e_g$ crystal-field splitting) may strongly affect magnetic properties via completely different mechanism. In strongly correlated materials even a small crystal-field splitting may result in electron localization on a  particular orbital.
Moreover, it turns out that the magnetic properties of a system strongly depend on  the particular orbitals at which electrons are localized. There are the so-called Goodenough-Kanamori-Anderson (GKA) rules\cite{Goodenough}, which describe the relation between the  orbital occuoation and the resulting magnetic coupling in systems with localized electrons. In describing these rules we will use the terminology of filled (two electrons), half-filled (a single electron) and empty orbitals, and will explain how these rules can be applied in most common geometries.

 It is easier to start with a direct overlap between $d$-orbitals ({\it direct exchange}), and then consider a more typical for TM compounds situation, when TM ions are separated by ligands, so that the corresponding $d$-orbitals practically  do not have a direct overlap with each other and all hopping processes occur via ligand $p$ orbitals (the so called {\it superexchange}).
\begin{figure}[t]
   \centering
  \includegraphics[width=0.49\textwidth]{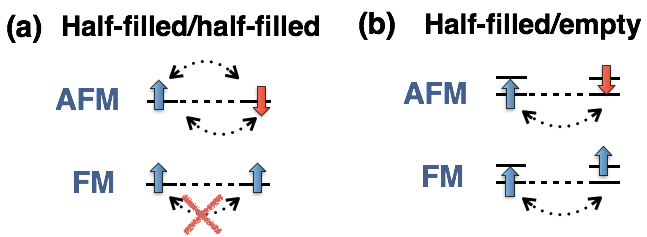}
  \caption{\label{fig:GKA} Sketch illustrating hopping processes in the case of a direct overlap between (a) two half-filled and (b) half-filled and empty $d$-orbitals.}
\end{figure}

{\it Direct exchange - case 1: The exchange coupling between two overlapping half-filled orbitals is strong and antiferromagnetic.} 

This situation is illustrated in Fig.~\ref{fig:GKA}(a). In the limit of large Hubbard repulsion, $U \gg t$, electrons are mostly localized on TM sites. If two electrons have different spin projections, i.e. are AFM coupled, they can sometimes hop from site to site and gain some kinetic energy. One may easily calculate a correction to the ground state energy due to this hopping, using second order perturbation theory with respect to $t/U$: $\delta E_{AFM}=\frac {2t^2}U$. Factor 2 appears here since both electrons can hop. $U$ is the energy of an intermediate perturbed state (when both electrons are on the same site) with respect to the ground state energy $E_0$. In the opposite situation of FM coupled spins electrons cannot hop due to Pauli principle and do not have this energy gain. Thus, the exchange integral is AFM (positive):
\begin{eqnarray}
\label{t2U}
J_{1} = E_{FM} - E_{AFM} = E_0 - \left(E_0 - \frac {2t^2}U \right)  = \frac {2t^2}U.
\end{eqnarray} 

{\it Direct exchange - case 2: The exchange coupling between overlapping half-filled and empty orbitals is week and FM.} 

First of all, since only one half-filled orbital can be directed along the line connecting two sites (otherwise there will be overlap between these two half-filled orbitals), only one electron can hop from site to site, and hence there will be no factor  2 in the expression for the exchange constant. Second, in this case Pauli principle does not restrict any hoppings, and both AFM and FM-coupled ions gain some energy due to these processes, Fig.~\ref{fig:GKA}(b). However one can see that this gain will be larger for FM, since the energy of the excited (virtual) state with two electrons on the same site in this case is smaller  - this state follows the Hund's rule, both electrons have the same spin and hence the energy of this state  is $U-J_H$, and not $U$ as it was for AFM. Corresponding exchange parameter is FM (negative):
\begin{eqnarray}
\label{t2JU2}
J_{2} = E_{FM} - E_{AFM} = E_0 - \frac {t^2}{U-J_H}  - E_0 + \frac {t^2}U \sim - \frac {t^2 J_H}{U^2}
\end{eqnarray}
(for $J_H<U$, which almost always is the case). This result can also be used for the case of overlap between (completely) filled and half-filled orbitals - one should just consider holes instead of electrons.

It is worthwhile to mention that for $3d$ TM ions $J_H \sim 1$ eV, while $U \sim 5-7$ eV\cite{khomskii2014transition}. 

Therefore, $|J_2|$ is usually (much) smaller than $J_{1}$, as defined in Eq.~\eqref{t2U}. This simple result has rather general implications. We see that in insulators the FM contributions to the exchange coupling are generally much smaller than the AFM ones: $J_1/|J_2| \sim U/J_H$.

This is the reason why most of the insulating TM compounds with localized electrons are AFM, not FM (in contrast to metals, which are typically FM). There must be special conditions, which allow FM $J_2$ to overcome $J_1$ (like small $U$, specific geometry or particular filling of $d$-levels which switches off the AFM contribution). Moreover, even if the  FM contribution dominates, the resulting Curie temperature is usually much smaller than the N\'eel temperature in AFM. Thus for example the  antiferro-orbital order (leading e.g. to overlap between half-filled and empty orbitals) does stabilize FM in YTiO$_3$, but T$_C \sim 30$ K, while in the AFM LaTiO$_3$ the ferro-orbital ordering(overlap of half-filled  with half-filled orbitals)  results in T$_N \sim 150$~K\cite{Hester1997,Cwik2003,Streltsov2005}. The Curie temperature in ferromagnetic NaCrGe$_2$O$_6$ is 6~K\cite{Vasiliev2005}, and in Ba$_2$NaOsO$_6$ T$_C \sim 7$~K\cite{Erickson2007}  -  much less than the typical values of Neel temperatures in TM oxides.
\begin{figure}[t]
   \centering
  \includegraphics[width=0.49\textwidth]{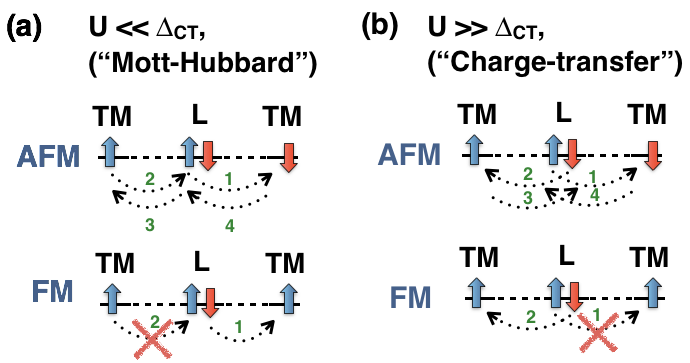}
  \caption{\label{fig:SU-t2} Schematic illustration of the hopping processes in superexchange between half-filled orbitals in two regimes (L stands for a ligand, TM is a transition metal ion).}
\end{figure}

One needs a  word of cation with respect to Eqs.~(\ref{t2U}-\ref{t2JU2}), which were derived for the case of only one electron and two electrons per site, \eqref{t2U} and \eqref{t2JU2} correspondingly. In real materials the occupation of $d$-states can be very different, and these formulas must be rewritten accordingly. One needs to calculate the energy of the intermediate state accurately. For example, in case of three electrons per site and half-filled/half-filled overlap between one of the orbitals $J_1 = 2t^2/(U+2J_H)$:  in initial state the hopping electron has Hund's rule ``attraction'' to the other two electrons at this site; this energy is lost in the virtual intermediate state when this electron is transferred to a neighbor. 
 
Also in writing down the expressions for different exchange constants we used the same value of $U$ for different orbital occupations. In fact this interaction is different for two electrons on the same ($U$) and on different orbitals ($U'$). In case of the $t_{2g}$ subshell one can make use the so-called Kanamori parameterization\cite{Kanamori1963}: $U' = U - 2J_H$; and in general one has to use the full atomic description, using Racah parameters $A,B$, and $C$)\cite{S.SuganoY.Tanabe1970}. This can change the exact expressions and numerical values of exchange constants, but the general qualitative rules formulated by Goodenough, Kanamori and Anderson (GKA rules) remain valid. 

Up to now we discussed direct overlap between $d$-orbitals. However, this situation is rather untypical for TM compounds, where TM ions are usually separated by ligands and are often quite far away from each other.   Since $d-d$ hopping scales as\cite{Andersen1978,Harrison1999} 
\begin{eqnarray}
\label{eq:dd-r-dependence}
t_{dd}  \sim r^{-(2l+1)}\sim r^{-5}, 
\end{eqnarray}
where $r$ is a distance between TM ions, the direct exchange is often rather inefficient. In this situation the electron hopping occurs via the ligand $p$ orbitals (superexchange). General rules about the overlap between filled, half-filled, and empty orbitals remain valid, but the analysis becomes more complicated, since one needs to take into account, in addition to $d$-states, also the energetics related to ligand orbital, and  all various exchange paths, which these orbitals provide.
\begin{figure}[t]
   \centering
  \includegraphics[width=0.49\textwidth]{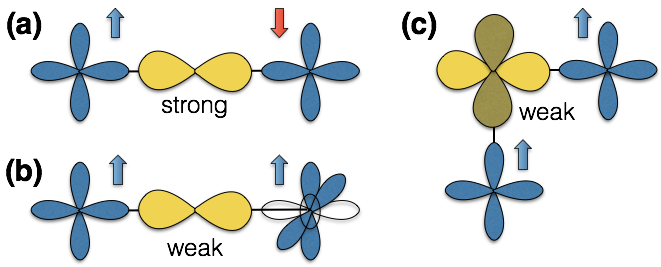}
  \caption{\label{fig:GKA2} Three main types of the superexchange interaction: (a) AFM superexchange between two half-filled $d$-orbitals via the \underline {same} $p$ orbital, Eq.~\eqref{t2U:SE}; (b) FM superexchange between half-filled and empty $d$-orbitals via the \underline {same} $p$ orbital, Eq.~\eqref{SE:hfe}; and (c) FM superexchange between two half-filled $d$-orbitals via \underline {different} $p$ orbitals, Eq.~\eqref{SE:hf-diff}. $d$orbitals of the TM ions are shown in blue (half-filled) and white (empty), while ligand $p$ orbitals are in yellow. In this figure only the $e_g$ orbitals are considered, corresponding plots for the $t_{2g}$ orbitals can be found, e.g. in \cite{Streltsov2008}.} 
\end{figure}

Here we will consider in details only the simplest situation of superexchange between two half-filled $d$-orbitals via $p$ orbital, as shown in Fig.~\ref{fig:GKA2}a (see \cite{khomskii2014transition} for a more complete analysis). In this case $d$-electrons will hop via ligand $p$ orbitals, corresponding hopping amplitudes are denoted as $t_{pd}$. There are two possibilities for this, as shown in Fig.~\ref{fig:SU-t2}. While the energy of the  excited state after the first such hop is the same,  $\Delta_{CT}$, the hopping processes on steps 2 and 3 are different. 

If Hubbard $U$ is smaller than the charge transfer energy $\Delta_{CT}$, $U \ll \Delta_{CT}$, then on the step 2 we move the $d$-electron on the vacant place in the $p$-shell (the energy of this state is $U$), and with the processes 3 and 4 we restore status quo. Corresponding  expression for the exchange constant reads as:
\begin{eqnarray}
\label{t2U:MH}
J \sim \frac {t_{pd}^4}{\Delta_{CT}U\Delta_{CT}} = \frac {\left(t^{eff}_{dd}\right)^2}{U}.
\end{eqnarray}
Here we intentionally introduced the effective $d-d$ hopping via $p$-orbitals, 
\begin{eqnarray}
t^{eff}_{dd} = t_{pd}^2/\Delta_{CT},
\end{eqnarray}
to demonstrate that the superexchange in this case has exactly the same form as a direct exchange defined in \eqref{t2U}.

In the opposite situation, $U \gg \Delta_{CT}$, it is easier to move at the second step the second electron from a ligand to another TM ion. In this case:
\begin{eqnarray}
\label{t2U:CT}
J \sim \frac {t_{pd}^4}{\Delta_{CT}(\Delta_{CT}+U_{pp}/2)\Delta_{CT}} = \frac {\left(t^{eff}_{dd}\right)^2}{\Delta_{CT}+U_{pp}/2}.
\end{eqnarray}
Here $U_{pp}$ is the on-site Coulomb repulsion on a $p$-shell of a ligand (it also takes into account intra-atomic exchange). 

The $U \ll \Delta_{CT}$ limit corresponds to Mott-Hubbard, while $U \gg \Delta_{CT}$ to charge-transfer insulators. However, in many real materials $U$ and $\Delta_{CT}$ can be of the same order, and one needs to take into account both contributions: 
\begin{eqnarray}
\label{t2U:SE}
J \sim \left(t^{eff}_{dd}\right)^2 \left( \frac 1U + \frac 1 {\Delta_{CT}+U_{pp}/2} \right).
\end{eqnarray}

\begin{figure}[t]
   \centering
  \includegraphics[width=0.49\textwidth]{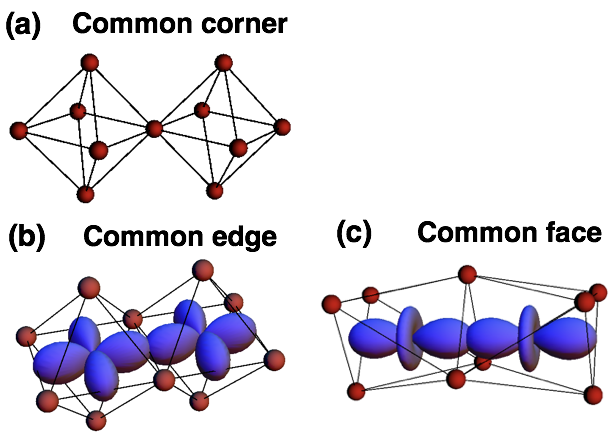}
\caption{\label{fig:packing} Three types of packing of the octahedra. For common edge and common face cases we also show $d$-orbitals with the largest direct overlap. Ligands are shown as brown balls.} 
\end{figure}

Without further details  we list  below the main contributions to the exchange interaction for three main geometries: when two neighboring MO$_6$ octahedra share their corners, edges and faces, see Fig.~\ref{fig:packing}.

\underline {Common corner}. Typical crystal structures with this geometries are: perovskites (normal, double, quadruple and layered). Since there is a ligand in between  TM ions, the dominating exchange is the 180$^{\circ}$ superexchange \eqref{t2U:SE}. The strongest will be the AFM exchange between half-filled $e_g$ orbitals via $\sigma$ ($p-d$) bond, Fig.~\ref{fig:GKA2}A. In addition, there can be also moderate AFM superexchange between half-filled $t_{2g}$ orbitals via the same $p$ orbital (described by the same \eqref{t2U:SE}), since $\pi$ bonding is much weaker than $\sigma$ ($t_{pd\sigma} \approx 2t_{pd\pi}$\cite{Harrison1999}). The last contribution   
is a weak FM exchange between half-filled and empty $d$-orbitals, Fig.~\ref{fig:GKA2}b:
\begin{eqnarray}
\label{SE:hfe}
J \sim -\left(t^{eff}_{dd}\right)^2 \left( \frac {J_H}{U^2} + \frac {J_H} {(\Delta_{CT}+U_{pp}/2)\Delta_{CT}} \right).
\end{eqnarray}

\underline {Common edge}. Typical materials with such crystal structures are: pyroxenes, delafossites, spinels (AM$_2$O$_4$), hexagonal ``213'' systems ((Li,Na)$_2$MO$_3$, see  Sec.~\ref{sec:molecules} and \ref{sec:Kitaev} for a detailed discussion).  There is a substantial direct $d-d$ overlap of two half-filled $t_{2g}$ orbitals (Fig.~\ref{fig:packing}b), which will result in a strong AFM exchange \eqref{t2U}. There will be also 90$^{\circ}$ superexchange interaction. First of all, a moderate AFM exchange appears via the same $p$ orbital, see Fig. ~\ref{fig:Kitaev} (one can use \eqref{t2U:SE} with appropriate choice of $t_{pd}$ in this case). Second, there also will be a FM superexchange between half-filled and empty $d$-orbitals, which is shown in Fig. 4 of \cite{Streltsov2008} and which can be described by \eqref{SE:hfe}. Finally, there is also a possible FM superexchange between two half-filled $e_{g}$ (or $t_{2g}$) orbitals via two \underline {different} $p$ orbitals as shown in Fig.~\ref{fig:GKA2}c (for the $t_{2g}$ orbitals see Fig.~5 of \cite{Streltsov2008}):
\begin{eqnarray}
\label{SE:hf-diff}
J \sim -  \frac {\left(t^{eff}_{dd}\right)^2 J_H^p} {(\Delta_{CT}+U_{pp}/2)\Delta_{CT}},
\end{eqnarray}
where $J_H^p$ stands for the Hund's exchange on the ligand site.

\underline {Common face}. Typical crystal structures: one dimensional or dimerized systems such as BaRuO$_3$, CsCuCl$_3$, or 6H-perovskites with general formula Ba$_3$(M1)(M2)$_2$O$_9$ (where M1 and M2 are metals) etc. The strongest exchange coupling is between the $a_{1g}$ orbitals ($a_{1g} = (xy+yz+zx)/\sqrt{3}$ in the local coordination system, where axes are directed towards ligands), see Fig.~\ref{fig:GKA2}c. Exceptionally large this contribution will be in case of $4d$ and $5d$ TM ions,  wave functions of which are more spatially extended than $3d$. This exchange is strong and AFM. It is interesting to note that the spin-orbital (Kugel-Khomskii) Hamiltonian, describing interplay between spin and orbital degrees of freedom, in this case has unusually high symmetry - SU(4)\cite{Kugel2015,Khomskii2016}.

In the end of this section we would like to mention that in principle there can be exchange processes via not one but several intermediate ions. Sometimes this exchange interaction is referred to as a super-super exchange\cite{DA2009,Reynaud2013,Markina2014}.

 \subsection{The double  exchange~\label{Sec:DE}}
Let us turn to the exchange interactions in metals. We consider not all metals, but only those in which local magnetic moments still exist. Moreover, we examine a situation, when there are two sets of electrons - one providing localized magnetic moments and another giving metallic conductivity. In  some sense this is an extreme situation, since in conventional metals the same electrons can be simultaneously mobile and provide magnetic moments. But in many materials, like manganites\cite{Tokura1999}, this is indeed a very good approximation: the part of the electrons are localized (due to strong Hubbard $U$), while the other (metallic) electrons can be added to a system, e.g. by  doping. One can assume that these two types of electrons interact with each other via intra-atomic Hund's exchange $J_H$:
\begin{eqnarray}
\label{eqn:DE}
H = -t \sum_{\langle ij \rangle \sigma} c^{\dagger} _{i \sigma} c_{j \sigma} - J_H \sum_i {\bf S}_i c^{\dagger} _{i \sigma} \pmb {\sigma} c_{i \sigma} + J\sum_{\langle ij \rangle}  {\bf S}_i {\bf S}_j.
\end{eqnarray}
Here $\pmb {\sigma}$ is the vector of Pauli matrices, while $\sigma$ is the spin. The first term gives a band spectrum of mobile electrons (described by operators $c^{\dagger} _{i \sigma}, c_{j \sigma}$), the second one introduces the coupling between mobile and localized electrons (with the spin ${\bf S}_i$). The last term is an exchange coupling between localized spins of neighboring sites.
\begin{figure}[t]
   \centering
  \includegraphics[width=0.49\textwidth]{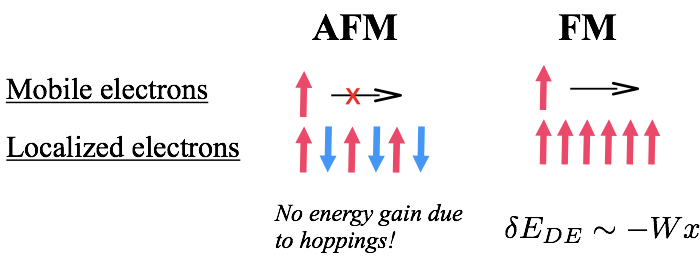}
  \caption{\label{fig:DE} Sketch illustrating the double exchange mechanism of ferromagnetism.}
\end{figure}

In the case of small doping all mobile electrons are at the bottom of a band, the width of which is defined by $t$ and the number of nearest neighbors $z$: $W \sim 2zt$. Thus, we can lower the total energy of a system considerably (by $\sim xW/2$, where $x$ is a concentration of mobile electrons), if mobile electrons would propagate through the lattice. However, if localized spins are AFM ordered, the intra-atomic Hund's exchange prevents (or at least strongly suppresses) such a propagation, since there are sites at which the spins of mobile and localized electrons would be  antiparallel, see Fig.~\ref{fig:DE}. Thus to gain kinetic energy of mobile electrons, it is better to make the system ferromagnetic.  We see that in contrast to the direct exchange  and superexchange, discussed in Sec.~\ref{sec:RealDesc} B and C, this mechanism, called in literature  {\it double exchange}, tends to stabilize ferromagnetism. Corresponding model, given by \eqref{eqn:DE} (sometimes omitting the last term,  with the assumption that $J_H$ is much larger than the other parameters of the system), is called the double exchange or ferromagnetic Kondo lattice model. Other details of this mechanism and the more detailed treatment of the model eqref{eqn:DE} can be found in the review \cite{Izyumov2001} and in original papers \cite{Zener1951,Anderson1955,Yosida1957,DeGennes1960,Kubo1972}. Here, we would like to mention just a couple of points. 

First of all, let us give some examples of systems, where the DE is operating. These are for example  manganites, such as La$_{1-x}$Sr$_x$MnO$_3$, where electrons in the narrow $t_{2g}$ band are considered as localized (and having local magnetic moments). By doping one may add some holes or electrons to the much wider $e_g$ band. The electrons or holes in the $e_g$ band play a role of itinerant carriers\cite{Zener1951}. Another example is CrO$_2$, where we do not need a doping to ``switch on'' the DE. There are localized electrons in the $xy$ band, which provide local magnetic moments, and itinerant electrons in the $xz/yz$ bands, which make the system ferromagnetic, hopping from site to site\cite{Korotin1998}.

Second, there can be a conventional direct or superexchange interaction between localized spins, described by the last term in \eqref{eqn:DE}, which is usually AFM as explained in Sec.~\ref{sec:RealDesc} B and C. The competition between the AFM superexchange and the FM double exchange can result in a canted magnetic state with the angle $\theta$ between neighboring spins  $\cos \frac {\theta} 2 \sim  tx/(4JS^2)$\cite{DeGennes1960} for appropriate concentration $x$, see also discussion in \cite{Nagaev1970,Kubo1972}.  Another more plausible option is that for small doping there may appear in a system, instead of homogeneous canting, a phase separation into the undoped antiferromagnetic matrix and the ferromagnetic droplets containing all doped electrons\cite{Kagan1999}. There are experimental indications that such phase separation indeed exists in low-doped manganites \cite{dagotto2003}.

Finally, there is an important question, what happens with the double exchange, if there appears a small band gap, which prevents propagation of itinerant electrons. Or, in other words, how the double exchange concept could be combined with the superexchange picture in a multiorbital case. While this is still a not completely solved problem, it was recently shown that the double exchange survives even in the insulating regime, if $J_H$ is large enough\cite{Nishimoto2012,Streltsov2016b}. Moreover, for a certain range of parameters there appears a phase with partially suppressed total magnetization. It is clear that the natural generalization of the double exchange model would  be a picture which would retain differentiation of electrons on more localized and more itinerant, but which does not require metallic conductivity. Obviously such a difference can be provided by a spatial ordering of corresponding orbitals. In Sec.~\ref{sec:OS} we discuss on the example of dimerized systems the interplay between the orbital-selective behaviour and magnetic properties, in particular the eventual suppression of double exchange by th formation of orbital-selective ``molecular'' states.

\subsection{The Jahn-Teller effect}
\begin{figure}[t]
   \centering
  \includegraphics[width=0.45\textwidth]{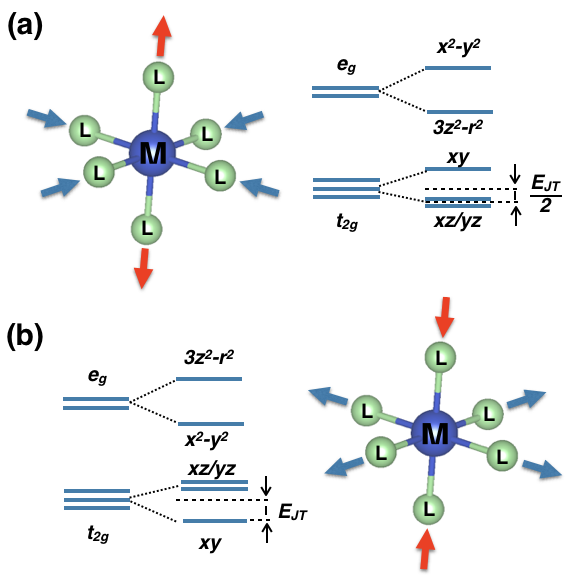}
  \caption{\label{fig:JT-elong} Tetragonal elongation (a) and compression (b) of a metal (M) - ligand (L) octahedron and corresponding splitting of the $d$-levels. In order to keep the volume of a crystal constant the elongation (compression) along one of the axes is accompanied by the compression (elongation) along two others.}
\end{figure}

Yet one more important factor, which we should mention here, is that for certain types of symmetry and for some electron occupations we can have an extra orbital degeneracy. This is the case, for example, for the TM ion having four $d$-electrons all with spin up (Mn$^{3+}$ or Cr$^{2+}$) in the octahedral coordination. Three electrons occupy the $t_{2g}$ levels, which are half-filled, and the fourth electron then goes to the $e_g$ state. But for regular octahedra these $e_g$ levels are doubly-degenerate. Thus, this extra electron can occupy any of these states: $3z^2 - r^2$,  or $x^2-y^2$, or any of their linear combination. This leads to the well-known instability, known as the Jahn-Teller (JT) effect: it is favorable to reduce the symmetry of a system, e.g. distorting the initially regular O$_6$ octahedron around TM in oxide, leading to the spitting of $d$-levels and to some gain in energy. Such splitting for the tetragonal elongation of O$_6$ octahedron is shown in Fig.~\ref{fig:JT-elong}(a). We see that such a distortion splits the $e_g$ levels, so that our fourth electron can now occupy the lowest $e_g$ level and can decrease its energy. This decrease turns out to be linear in distortion $u$,  i.e. $\delta E_{kin} \sim -gu$,  as the level splitting in a Zeeman effect. Here $g$ is a parameter characterizing the coupling between an electronic subsystem and a lattice, $u$ is a deformation. Of course this distortion leads to an elastic energy loss, which, however, is only quadratic in the displacement,  $\delta E_{elast} \sim Bu^2/2$ ($B$ is elastic modulus). The linear electronic energy gain always wins, and the minimum energy will be reached for finite distortion, in this case $u=\pm g/B$. This is, in simple terms, the essence of the JT theorem (which, according to Teller himself, was first suggested to him by Landau, see. App. A.2 in \cite{khomskii2014transition}).

For  isolated centers the Jahn-Teller instability leads to very interesting quantum effects, including the geometric (Berry) phase (which actually first appeared in the literature just in this context\cite{Longuet-Higgins1}, long before the famous works of Sir M. Berry). But for us it is more important that for concentrated solids one can get in this situation structural phase transitions with corresponding orbital ordering – see e.g. \cite{KK-UFN}.  Moreover, it is not clear what comes first - the JT distortions and then orbitals follow, or vice versa. Indeed, in addition to the electron-phonon mechanism of the JT effect described above, there is another one, the  so-called superexchange  (or Kugel-Khomskii) mechanism\cite{KK-UFN}. We already know that a system may gain an exchange energy by setting up some orbital ordering (e.g. occupying the overlapping half-filled orbitals we gain the energy proportional to \eqref{t2U:SE}), and the crystal lattice will react on this by corresponding (JT) distortions. In fact instead of a real orbital ordering in an undistorted high temperature phase one should rather speak about short-range orbital correlations.

Band structure calculations show that there can appear an orbital ordering even in the absence of the JT distortions, just due to the superexchange mechanism, and if we then allow for the lattice relaxation, lattice will relax to the JT distorted structure (in the LDA+U method\cite{Liechtenstein1995}\footnote{LDA - local density approximation.}, i.e. including electronic correlations described by the Hubbard's $U$, which are needed to localize electrons on particular orbitals)\cite{Streltsov12Cu,Streltsov2014d}. The more sophisticated LDA+DMFT calculations\footnote{DMFT - dynamical mean-field theory}, however, show that both the electron-phonon and the superexchange mechanisms are important, and they together determine the temperature of the JT transition\cite{Pavarini2008,Leonov08,Pavarini2010}. We will not discuss this big and very interesting field here; but in dealing with real systems with orbital degeneracy one always has to keep in mind the possibility of the JT distortions, which could result in the formation of an orbital ordering.

\subsection{The spin-orbit coupling\label{sec:SOC-intro}}
When dealing with the TM compounds, especially with $4d$ and  $5d$ TM, one has to also take into account the real (relativistic) spin-orbit coupling (SOC). It becomes large, comparable to many other parameters, especially for $5d$ compounds. Still usually the spin-orbit constant $\lambda$  ($\sim$0.5 eV for such ions as Ir, Pt) is smaller than the $t_{2g} - e_g$ crystal field splitting $\Delta_{CFS}$ ($=10Dq$), which for $5d$ oxides is typically $\sim$3-4 eV.  For the $e_g$ electrons the crystal field quenches the orbital moment and  the SOC. Therefore we should only expect strong effects of the SOC for systems with partially-filled  $t_{2g}$ subshells. But to these actually belong most of $4d$ and $5d$ compounds, since $4d$  and $5d$ TM ions are usually in the low-spin state, see Sec.~\ref{sec:CFS}.

For the $t_{2g}$ subshell with triply-degenerate orbitals one can, applying the Wigner-Eckart theorem, describe orbitals using the equivalent orbital moment  $l_{eff} = 1$.  
Indeed, matrix elements of orbital moment for three $t_{2g}$ orbitals coincide with those for $l=1$ up to a sign of spin-orbit constant\cite{Abragam}.  In the following we will use this very convenient language for description of $4d$ and $5d$ orbitals and for simplicityoften omit the  ``eff'' subscript. One has only to take care of the magnitude and the sign of the effective spin-orbit coupling $\lambda_{eff}$, when written for this effective moment.

Two remarks have to be made right away. The first one is that when we include the SOC, the electron-hole symmetry existing for $t_{2g}$ shell is lost. Without SOC the properties of systems containing  one and five $t_{2g}$ electrons, and also two and four of them are equivalent, with the electron--hole substitution. Therefore one can easily ``translate'' the results obtained for example for one electron to those with five  electrons (or one hole) in $t_{2g}$ subshell. This, however, is no more the case in the presence of (strong) SOC.
\begin{figure}[t]
   \centering
  \includegraphics[width=0.49\textwidth]{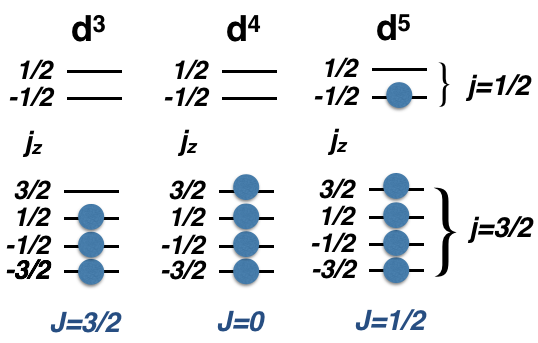}
  \caption{\label{jj-d4-d5} The $jj$ scheme for $d^3$, $d^4$, and $d^5$ configurations (it is assumed that the $t_{2g}-e_g$ crystal-field splitting is large, so that all electrons are on the $t_{2g}$ levels).}
\end{figure}

The second point is the way we consider the SOC in many-electron atoms or ions. In principle it is a complicated many-particle problem.  The detailed  analysis of the structure of atomic terms, with real atomic parameters (Racah parameters $A$, $B$, and $C$, or intra-orbital and inter-orbital Hubbard repulsions $U$ and $U’$ and the Hund’s interaction $J_H$) is described for example in \cite{S.SuganoY.Tanabe1970,Abragam}.  Generally,  in atomic physics one usually considers two limits, or two approximations. From Dirac equation one gets the SOC for one electron, $\zeta l_i s_i$\cite{LandauQ}, with the positive coupling constant $\zeta$ (and dependent on the atomic number, see below). For many-electron atoms or ions with relatively weak SOC (weaker than the Hund’s rule intra-atomic exchange) one usually uses the $LS$, or Russel-Saunders approximation. In this one, according to the first Hund’s rule, see e.g.\cite{khomskii2014transition}, one first forms the total spin  $\bf S = \sum_i  \bf s_i$, and the total orbital  moment $\bf L = \sum_i \bf l_i$, and then uses the spin-orbit interaction for these total moments
\begin{eqnarray}
\label{eq:full-SOC}
H_{SOC} = \lambda \mathbf L  \mathbf S.                                                                       
\end{eqnarray}
The energy contribution due to the SOC can be expressed via the total moment $\bf J$, defined as $\mathbf J =  \mathbf L  + \mathbf S$:
\begin{eqnarray}
E_{SOC} = \langle \lambda \mathbf L  \mathbf S \rangle = \frac {\lambda}2 
\left( J(J+1) - L(L+1) - S(S+1) \right), \nonumber
\end{eqnarray}
since $\mathbf J^2 = \mathbf L^2 + \mathbf S^2  + 2 \mathbf L \mathbf S$. The SOC constant then is $\lambda = \pm \zeta/2S$, where one takes plus for the less-than-half-filled shells and minus for th more-than-half-filled shells. This finally leads to the second (or third) Hund’s rule: for the less-than-half-filled shells ($\lambda >0$) we have a normal order of multiplets (the terms with the smaller $J$ have lower $E_{SOC}$), and the “inverted” multiplet  order (the lowest multiplets are those  with the maximum $J$) for the more-than-half-filled shells.

When dealing  with the effective moment $l=1$ and the effective SOC for the $t_{2g}$ shell, the sign of $\lambda_{eff}$ turns out to be opposite \cite{Abragam,khomskii2014transition}, so that we have a reversed multiplet order: multiplets with the larger $J$ lie lower in energy for the less-than-half-filled $t_{2g}$ shells,  and we have a normal order for the more-than-half-filled shells. It is this factor that finally gives an electron-hole asymmetry for this case. Thus according to these rules for a $d^1$ configuration, with $L=1$ and $S=1/2$,  the possible values of the  total moment are $J=1/2$ and $J=3/2$, and according to the rules formulated above the lowest multiplet is the quartet $J=3/2$. However for five $d$-electrons (one hole in the $t_{2g}$ shell) the multiplet order will be inverted, so that the ground state of such ion would be a doublet $J=1/2$. This is the state often invoked nowadays for the compounds containing Ir$^{4+}$ ($t_{2g}^5$), see discussion  below.
\begin{figure}[t]
   \centering
  \includegraphics[width=0.49\textwidth]{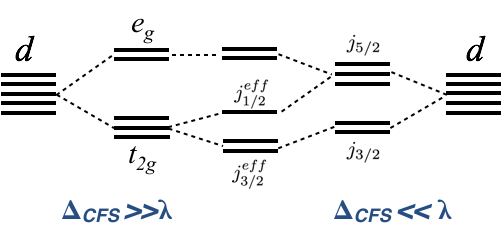}
  \caption{\label{CFS-SOC} Scheme, illustrating level splitting in the presence of  cubic crystal-filed (CF) and of spin-orbit coupling.}
\end{figure}

This treatment is applicable for light elements, with relatively weak SOC. In the opposite limit of very strong SOC, realized for example for rare earths of for actinides, one usually uses another approximation – the so called $jj$ coupling scheme  
(realized if the SOC constant $\lambda$ is larger than the Hund's exchange  $J_H$). 
In this scheme one first couples for each electron its spin and angular moments to the total moment of an electron,
\begin{eqnarray}
\mathbf {j}_i = \mathbf {l}_i + \mathbf {s}_i,
\end{eqnarray}
and then one forms the total moment out of those moments for individual electrons:
\begin{eqnarray}
\mathbf  J=\sum_i \mathbf {j}_i,
\end{eqnarray}
and then forms a total moment $J$ out of these $j$'s for indvidual electrons. In this scheme a strong SOC splits the one-electron $d$-state into $j=5/2$ and $j=3/2$,  and then the other interactions may lift the degeneracy of these levels. Note, that by this we violate the first Hund's rule, first of all taking care of the spin-orbit coupling (assumed to be stronger than the Hund's exchange). A general scheme of the $d$-levels splittings in the presence of crystal-field and SOC is shown in Fig.~\ref{CFS-SOC}.

The $3d$ compounds are definitely better described by the $LS$ (Russel-Saunders) coupling scheme, and probably so are the most of $4d$ systems. But with the $5d$ materials the situation is not so clear. It might be that they are already ``in between'' the $LS$ and $jj$ couplings.

For some $d$ counts these two pictures give qualitatively similar results, but for some others the conclusions might be different. Thus, for example, for the low-spin $d^4$ configuration in the $LS$ scheme $L=1, S=1$, and the ground state should be $J=0$ singlet. The same conclusion would one get in the $jj$ scheme. In this scheme we have single-particle states in the form of low-lying $j=3/2$ quartet and higher-lying $j=1/2$ doublet, see Fig.~\ref{jj-d4-d5}. Four $d$-electrons would then occupy all states of the 3/2 quartet, i.e. the total $J$ would be also zero.

The same is true for the most widely discussed case of $d^5$ occupation, as in Ir$^{4+}$. In the $LS$ coupling scheme, as mentioned above, we would have $L=1, S=1/2$, and the ground state woud be a Kramers doublet $J=1/2$.  In the $jj$ scheme we should fill the levels shown in Fig.~\ref{jj-d4-d5} by five electrons, which would completely fill the lowest quartet, and the fifth electron will be in the $j=1/2$ doublet, as in the $LS$ scheme. But for example the situation would be different for $d^3$ occupation.  In the $LS$ scheme these three electrons would fill all $t_{2g}$ levels  (the high spin state), so that the net orbital moment would be $L=0$, and what remains would be a pure spin $S=3/2$  state, without any influence of the SOC. In the $jj$ scheme we also would have three electrons on a quartet, but not a quartet  $S=3/2$, but $j=3/2$ quartet, Fig.~\ref{jj-d4-d5}. Consequently the form of corresponding wave functions, the values of $g$ factors etc., would  be different, see e.g.~\cite{Matsuura2013}. Very recently these effects were indeed observed for $5d^3$ systems Ca$_3$LiOsO$_6$ and Ba$_2$YOsO$_6$\cite{Taylor2017}.

It is also worth mentioning that  all the band structure calculations based on the density functional theory (DFT)\cite{Kohn1999} are in fact dealing with one-electron states (one Slater determinant). In this sense they describe the SOC in the $jj$ scheme, which also operates with one-electron states, before combining them into a total $J$ state. Also experimentalists very often use the description with the energy schemes similar to Fig.~\ref{jj-d4-d5}. One has to realize though that the real atomic terms, real multiplets are many-particle states, especially in the $LS$ coupling scheme.

One more comment is in place here. We have said above that the SOC becomes stronger with increasing atomic number of an element $Z$, and because of that the heavier elements like $5d$ TM’s may be already close to the $jj$ coupling scheme. Most often in the literature one gives the estimate that the spin-orbit coupling constant  $\lambda \sim Z^4$, where $Z$ is the atomic number of an element; this became already an accepted notion. But in the famous textbook \cite{LandauQ} it is shown that in fact this relation should rather be $\lambda \sim Z^2$, not $Z^4$ \cite{LandauQ}! And indeed a comparison with the experimental data show that this estimate is much closer to reality (though both are of course the ``order-of-magnitude estimates''). For example compare Ir and V.  Ir has atomic number $Z=77$ and $\lambda = 400 $ meV\cite{Friedel1969}. V has atomic number $Z=23$ and $\lambda \sim 30 $ meV\cite{Abragam}. In effect $\lambda_{Ir}/\lambda_V=13.3$. Now, the ``Landau estimate'' gives $(Z_{Ir}/Z_V)^2  = 11.2$, but the more commonly used ``rule'' $\lambda \sim Z^4$ would give $(Z_{Ir}/Z_V)^2  = 125$  -- way off!  Thus, it seems that the dependence $\lambda \sim Z^2$ is indeed a correct one.

\begin{table*}[t!]
\centering \caption{\label{Tab:red-dim} Examples of materials with effective reduction of the dimensionality due to orbital degrees freedom.}
\vspace{0.2cm}
\begin{tabular}{llcccc}
\hline
\hline
                   &{\bf Type of reduction} & {\bf Materials}                        & {\bf References}\\
\hline
1D$\to$0D& zigzag chains $\to$ $S=0$ dimers & NaTiSi$_2$O$_6$    & \cite{Wezel2006,Streltsov2008}\\
1D$\to$0D& chains $\to$ dimers  & TiOCl                           &\cite{Seidel2003} \\
\hline
2D$\to$0D& triangular layers $\to$ isolated triangles ($S=0$)  & LiVO$_2$                    &  \cite{Kobayashi1969,Katayama2009}\\
\hline
3D$\to$0D& Spinel $\to$ $S=0$ heptamers  & AlV$_2$O$_4$      &  \cite{Horibe2006,Uehara2015}\\
3D$\to$0D& Spinel $\to$ $S=0$ octamers  & CuIr$_2$S$_4$           & \cite{Radaelli2002,Khomskii2005a} \\
3D$\to$1D& Spinel $\to$ tetramerized chains ($S=0$)  & MgTi$_2$O$_4$         &  \cite{Schmidt2004,Khomskii2005a} \\
3D$\to$1D& 3D Perovskite $\to$ AFM $S=1/2$ chains  & KCuF$_3$              &  \cite{KK-JETP,Satija1980} \\
3D$\to$1D& 3D Pyrochlore $\to$ Haldane chains & Tl$_2$Ru$_2$O$_7$      & \cite{Lee2006} \\
\hline
\hline
\end{tabular}
\end{table*}

 \section{Effective reduction of the dimensionality due to orbital degrees of freedom and its consequences\label{sec:red-dim}}
The original investigations of ``orbital physics'' in solids were mostly concentrated on the study of the effects connected with orbital degeneracy and with the resulting phase transitions - the cooperative JT effect, or orbital ordering (these terms actually denote the same phenomenon, just stressing different aspects of it). These effects were discussed in many books and review articles, e.g. \cite{Gehring1975,KK-UFN,Tokura2000}. Lately some novel aspects of orbital physics  attracted significant attention and came to the forefront. In the present review we will mostly concentrate  on this novel development; the older more ``classical'' parts of this field one can find in the literature cited above.
					
We start by discussing the phenomenon which was highlighted   relatively recently  and which was shown to lead to many interesting consequences. This is the reduction of  the effective dimensionality of electronic and magnetic subsystems, which is the result of directional character of $d$-orbitals, see e.g. Fig.~\ref{fig:cubic-orbitals}. We will describe these effects on several examples, before formulating general conclusions.  In Tab.~\ref{Tab:red-dim} we give a list (far from compete!) of several materials in which the phenomenon of reduction of effective dimensionality was observed experimentally.

\subsection{Formation of low-dimensional magnetic systems due to orbital ordering}

\begin{figure}[t]
   \centering
  \includegraphics[width=0.49\textwidth]{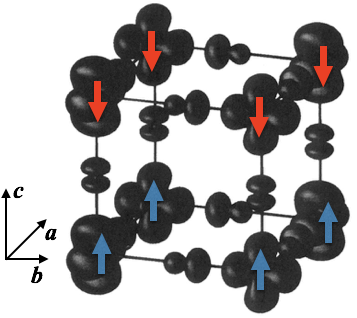}
  \caption{\label{fig:KCuF3} Spin density distribution (difference between charge density for spin up and down) obtained in the LDA+U calculations for KCuF$_3$\cite{Liechtenstein1995}. Cu ions are in the corners of the cube, F ions are in the middle of its edges. One may see that a single hole is localized on the $x^2-z^2$ and $y^2-z^2$ orbitals. This results in the antiferromagnetism of the A-type shown by arrows.}
\end{figure}

The simplest example, known already long ago, is the formation of low-dimensional magnetic systems in materials which just by crystal structure are the usual three-dimensional ones. Probably the most striking example is KCuF$_3$. This is an insulating perovskite, with basically cubic lattice, containing classical JT ions Cu$^{2+}$ ($t_{2g}^6 e_g^3$), with one hole in doubly-degenerate $e_g$ orbitals. Due to electron-lattice (JT) interaction\cite{Gehring1975} and superexchange mechanism\cite{KK-JETP,KK-UFN} there occurs in KCuF$_3$ an orbital ordering with the (half-filled) hole orbitals shown in Fig.~\ref{fig:KCuF3}.

Remembering the GKA rules, discussed in Sec.~\ref{sec:GKA}, we expect that there should exist in this system a strong  antiferromagnetic exchange along the $c$ direction, in which these orbitals strongly overlap (via corresponding $p$ orbitals of F). The coupling in the $ab$ plane is  ferromagnetic and much weaker: the half-filled orbitals here are orthogonal to each other, and there is the overlap only between half-filled and completely filled (in electron picture) or between half-filled and empty (in hole representation) orbitals. And indeed magnetic properties of KCuF$_3$ ideally correspond to these expectations: this system turns out to be a quasi-one dimensional antiferromagnet, with weak ferromagnetic coupling between these AFM chains, which finally leads to the long-range magnetic ordering of A-type (FM layers stacked antiferromagnetically, see Fig. \ref{fig:KCuF3}). And in effect this material, which is crystallographically cubic, turns out to be magnetically one of the best 1D antiferromagnet known\cite{Satija1980}! And this is completely due to corresponding  orbital ordering, with the resulting strongly anisotropic electron hopping and exchange interaction.

There exist other materials  in which orbital ordering leads to the formation of low-dimensional magnetic systems. A rather striking example is provided by pyrochlore Tl$_2$Ru$_2$O$_7$. In this, also crystallographically $3D$ cubic material, with Ru$^{4+}$ ($S=1$) there appears below phase transition at T$_c=$120 K a state with the spin gap. However structural studies did not show any apparent distortion which could have lead to the formation of singlet dimers, etc. The explanation proposed \cite{Lee2006} is that the orbital ordering appearing in Tl$_2$Ru$_2$O$_7$ below T$_c$ leads to the formation of magnetically quasi-one dimensional structures, chains of $S=1$ ions. Such objects - chains with integer spin are very well known in ``spin science’’ and they are called Haldane chains. In contrast to half-integer spin chains they exhibit a gap (a spin gap) in the spin excitation spectrum~\cite{Haldane-83,Affleck1989} and topologically-protected edge (here end) states.

\subsection{``1D-zation'' of electron spectrum and orbitally-driven Peierls state\label{sec:orb-Peierls}}

There exist other materials with similar reduction of effective dimensionality of magnetic subsystem. But even more drastic consequences could result from the reduction of dimensionality in the electronic subsystem. This is often related to the special properties of the low-dimensional, especially one-dimensional systems, in particular to the tendency of such systems to experience Peierls-like distortion.

In the Tab.~\ref{Tab:red-dim} we list some materials in which orbital structure leads to the reduction of effective dimensionality of electronic subsystem, in particular resulting in the formation of a Peierls-like state. Of course we cannot describe in this review all these examples; we concentrate on the most representative (and easy to explain) cases.

\begin{figure}[t]
   \centering
  \includegraphics[width=0.4\textwidth]{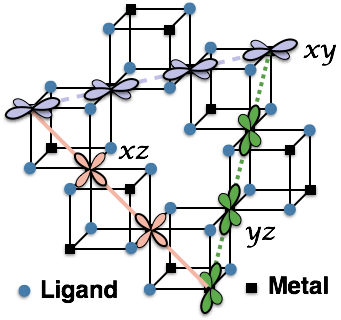}
\caption{\label{fig:spinel} Formation of 1D chains in spinels due to orbital degrees of freedom. Transition metal ions (squares) are at the B-sites of AB$_2$L$_4$ spinels. Ligands (L) are shown by blue circles.}
\end{figure}

\begin{figure}[t]
   \centering
  \includegraphics[width=0.49\textwidth]{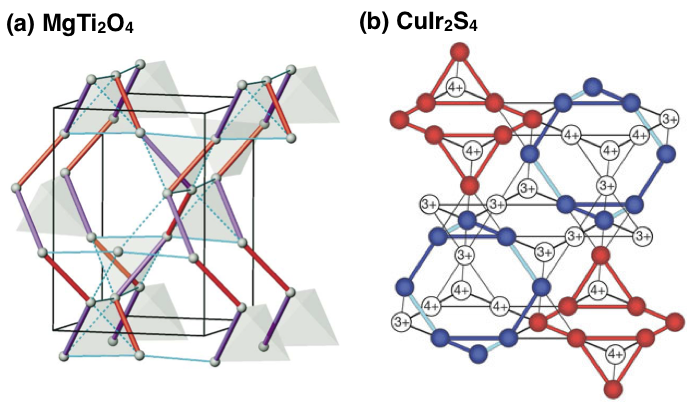}
\caption{\label{fig:MgTiCuIr} Crystal structure of MgTi$_2$O$_4$ and CuIr$_2$S$_4$ in the low temperature distorted phase (taken from \cite{Schmidt2004,Radaelli2002}). Transition metal ions are shown by circles. Different bond lengths are shown by different color.}
\end{figure}

Probably the most spectacular example is the formation of exotic superstructure in MgTi$_2$O$_4$ (spirals)\cite{Schmidt2004} and in CuIr$_2$S$_4$ (octamers)\cite{Radaelli2002}. These materials are spinels with TM in B-sites, Fig.~\ref{fig:spinel}. In both there occurs a structural transition from the cubic to tetragonal phase with decreasing temperature. But, besides that, there appears in these systems extra distortions, leading to the formation of beautiful superstructures. Short and long Ti-Ti bonds, forming strange ``spirals'' are formed in the low temperature phase of MgTi$_2$O$_4$  (see Fig.~\ref{fig:MgTiCuIr}a). Even more nontrivial superstructure was found by the same group in CuIr$_2$S$_4$: there occurs in this system below 230 K a charge ordering of Ir ions (the average valence Ir$^{3.5+}$) into Ir$^{4+}$($t_{2g}^5$) and nonmagnetic (low-spin) Ir$^{3+}$($t_{2g}^6$), and these species form beautiful octamers, see Fig.~\ref{fig:MgTiCuIr}b. Besides that, in Ir$^{4+}$ octamers there occurs extra dimerization, with the formation of short Ir$^{4+}$-Ir$^{4+}$ singlet dimers, which makes the whole material nonmagnetic.

In the original publications\cite{Schmidt2004,Radaelli2002} there was no explanation of the mechanism  of the formation of these superstructures. But one can find a very straightforward explanation of the observed superstructures if one takes into account orbital dependence of electronic structure in these spinels\cite{Khomskii2005a}.

In both cases we are dealing with the systems with partially-filled $t_{2g}$ levels. As one can see from Fig.~\ref{fig:spinel}, in the geometry of B-sites of a spinel lattice, with edge-sharing TiO$_6$ or IrS$_6$ octahedra, there occurs strong direct overlap of particular $t_{2g}$ orbitals in a particular direction. Thus e.g. the $xy$ orbital of one site strongly overlaps with similar $xy$ orbital along $xy$ direction, but not with the two other orbitals. Similarly, $yz$ orbitals  overlap and have strong hopping to the same $yz$ orbital in the $yz$ direction. Now, the structure of B-sites of a spinel can be visualized as consisting of straight chains running in the $xy$, $xz$ and $yz$ directions. It may seem just an artificial construction, but just for $t_{2g}$ orbitals it acquires real significance. We see that, hopping from site to site, the electrons for example in the $xy$ orbital would remain on the same orbital in corresponding $xy$ chain, and similarly for the $xz$ and $yz$ orbitals. In effect, if we only include direct $d-d$ overlap and hopping, electronic structure of these, basically cubic materials, would be composed of three one-dimensional bands, $xy$, $xz$ and $yz$.

Now, the famous Peierls effect tells us  that the metallic state of such one-dimensional systems is unstable towards the formation of superstructure which opens a gap at the Fermi-level (see Sec.~\ref{sec:basic-concept}). For half-filled bands it would lead to dimerization - the best-known case. But actually the same instability exists also for other band fillings. Thus, for quarter-filled bands we would get tetramerization,  for 1/3 filled band --- trimerization, etc.  And this was the explanation proposed in \cite{Khomskii2005a} for superstructures observed in MgTi$_2$O$_4$ and CuIr$_2$S$_4$. One can easily see that in both these cases we would have 1/4-filled bands: doubly-degenerate $xz$ and $yz$ bands in MgTi$_2$O$_4$ and 1/4 (or rather 3/4) filled band in CuIr$_2$S$_4$. And the exotic and puzzling superstructures observed in \cite{Schmidt2004,Radaelli2002} found natural explanation if one only takes ``right point of view'' and looks at what happens in 1D bands determining the electronic structure of these materials. In both these cases we have a simple tetramerization on the straight chains: in the $xz$ and $yz$ chains in MgTi$_2$O$_4$, and in all directions in CuIr$_2$S$_4$. Thus, this, rather strongly simplified picture (we ignored for example a possible electron hopping  via ligands - oxygen, sulphur), gives a natural explanation of very exotic and beautiful superstructures found in MgTi$_2$O$_4$ and CuIr$_2$S$_4$\cite{Khomskii2005a}.

There exist also other materials of the same class in which this physics can be in action. For example such  can  be the situation in $V$ spinels, like ZnV$_2$O$_4$\cite{Reehuis2003,Lee2004}. This material has caused quite a discussion in theoretical community, several models were proposed to explain the superstructures observed in it \cite{Tsunetsugu2003,Tchernyshyov2004,Valenti2007,Pardo2008,Streltsov2015MISM}. The final explanation of the properties of this system is still not agreed upon;  but in any case all proposed pictures were based on the important role of orbital degrees of freedom in determining its properties.

\subsection{Novel states close to Mott transition: ``Molecules'' in solids\label{sec:molecules}}

In the previous sections we have seen that there may appear in solids with correlated electrons some clusters, e.g. dimers, in which electrons behave as practically delocalized, whereas there is still rather weak hopping between such clusters. One can often describe such clusters using the treatment developed for molecules. In concentrated solids such objects can appear when the whole system is relatively close to localized-itinerant crossover, i.e. close to Mott transition.
\begin{figure}[t]
   \centering
  \includegraphics[width=0.49\textwidth]{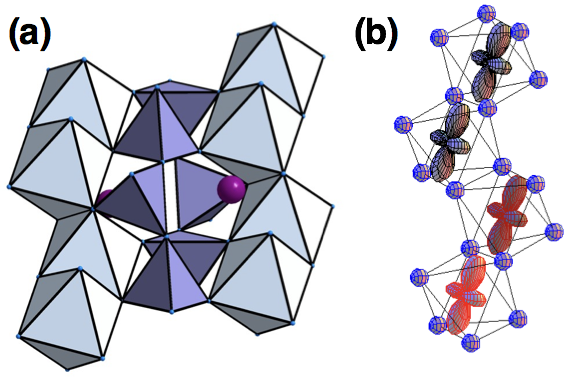}
\caption{\label{fig:pyroxenes} (a) Crystal structure of pyroxenes. Metal are inside octahedra, which form zig-zag chains. (b) Orbital ordering, which is stabilized in the low-temperature phase of NaTiSi$_2$O$_6$.}
\end{figure}

Usually, when thinking about Mott transitions, one has in mind the situation when on one side of the transition we have a homogeneous Mott insulator, and on the other side we are dealing with a homogeneous metallic state described for example by the Fermi-liquid theory. However, the experience collected in the last years demonstrated that this is not the only possible situation. It turns out that in many real systems electron delocalization first occurs in finite clusters - dimers, trimers, or sometimes larger clusters, whereas between those we still have weak hopping and the whole system still behaves as an insulator. And only at a later stage, e.g. at still higher pressures, the whole material may become metallic. 

In order to understand whether a system is in such state, one can compare the metal-metal distances in a compound under consideration with those met in pure metals, $D_{met}$, see Tab.~\ref{Tab:metal-metal}. If some distances are smaller than $D_{met}$, than this can be a signature of the formation of ``molecules'' in a given system.

The first example of formation of such ``molecules'' in bulk solids due to a particular orbital ordering is pyroxene NaTiSi$_2$O$_6$. Pyroxenes are a big class of materials which are yet not very popular among physicists, but which are extremely important in geology: these are silicates, one of the main rock-forming minerals. They constitute up to 20\% 
of the Earth's crust and are important constituents of the upper mantle\cite{anderson2007}.
\begin{table}
\centering \caption{\label{Tab:metal-metal} Metal-metal bonds, $D_{met}$, in pure metals. The distances are given in~\AA.}
\vspace{0.2cm}
\begin{tabular}{lllllllll}
\hline
\hline
{\bf $3d$:} &{\bf Ti} & {\bf V} & {\bf Cr} & {\bf Mn} & {\bf Fe} &  {\bf Co} & {\bf Ni} & {\bf Cu} \\
             & 2.896    & 2.622   & 2.498    & 2.734    & 2.482    &  2.506    & 2.492    & 2.556  \\
\hline
{\bf $4d$:} & {\bf Zr} & {\bf Nb} & {\bf Mo} & {\bf Tc} & {\bf Ru} &  {\bf Rh} & {\bf Pd} & {\bf Ag} \\
             & 3.180  & 2.858     & 2.726    & -           &  2.650   & 2.690     & 2.752    &  2.890 \\
\hline
{\bf $5d$:} & {\bf Hf} & {\bf Ta} & {\bf W} & {\bf Re} & {\bf Os} &  {\bf Ir} & {\bf Pt} & {\bf Au} \\
                  & 3.128  & 2.860     & 2.740   & 2.742   &  2.676    & 2.714     & 2.746    &  2.884 \\
\hline
\hline
\end{tabular}
\end{table}

These are quasi-one-dimensional compounds, containing zigzag chains of MO$_6$ octahedra sharing common edges, and in between there are SiO$_4$ (or GeO$_4$) tetrahedra, see Fig.~\ref{fig:pyroxenes}(a).  Material we want to discuss is NaTiSi$_2$O$_6$, with Ti$^{3+}$ ($d^1$). It is paramagnetic, with the susceptibility at high temperatures following Bonner-Fisher curve for a one-dimensional antiferromagnet with $S=1/2$. However this behaviour is interrupted at T$_c$ = 210 K, below which it is practically diamagnetic\cite{Isobe2002}.

Ab-initio calculations demonstrated that, whereas at high temperatures one $d$-electron of Ti occupies more or less equally all three $t_{2g}$ states, below T$_c$ there occurs ferro-orbital ordering, with occupied orbitals shown in Fig.~\ref{fig:pyroxenes}(b)\cite{Streltsov2008}. We see that after such ordering the system is practically divided into  dimers, weakly connected with each other: the exchange coupling inside such dimers is strongly antiferromagnetic, $\sim 400$ K, whereas the exchange between dimers is close to zero, and most probably is weakly ferromagnetic\cite{Streltsov2008}. In effect the material which was a one-dimensional antiferromagnet above T$_c$, becomes split below T$_c$ into singlet dimers. And this is predominantly due to particular orbital ordering; one even does not have to move ions (but of course in reality also the Ti-Ti distances inside and  between these dimers become different). This is a very clear example of reduced dimensionality and formation of singlet  ``molecules'' due to directional character of orbitals and due to a particular type of orbital ordering.
\begin{figure}[t]
   \centering
  \includegraphics[width=0.49\textwidth]{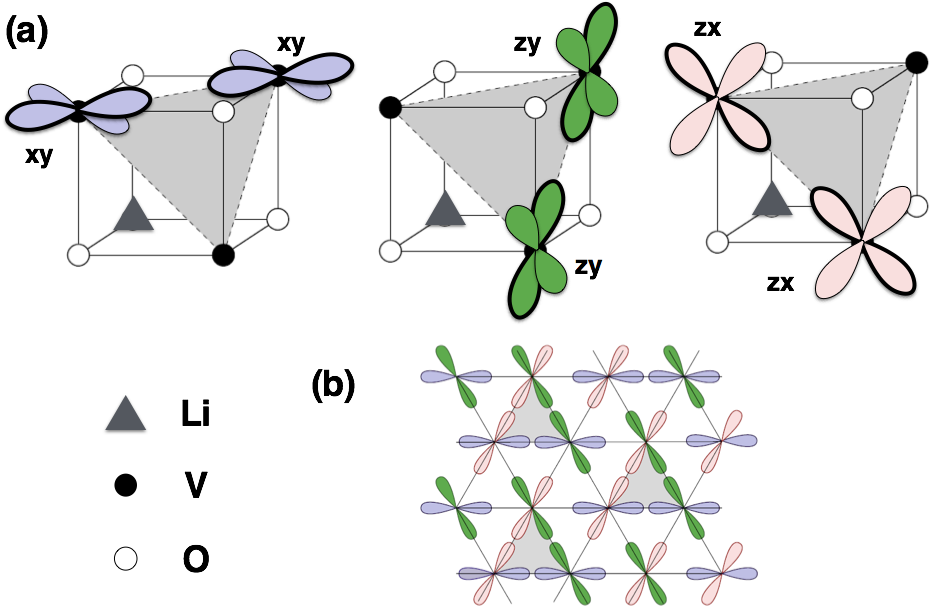}
\caption{\label{fig:LiVO2} (a) Orientation of $t_{2g}$ orbitals at $B$-sites of a spinel lattice. Out of three  $t_{2g}$ orbitals there are two for each V site which have a direct overlap with neighboring V ions  in LiVO$_2$.  Two ``active'' lobes (bold) of any of these orbital responsible for this overlap lie in the triangular layer of V, while two other lobes are perpendicular to this plane. (b) Orbital ordering (only ``active'' lobes are shown), which results in trimerization in LiVO$_2$.}
\end{figure}

Another such example is LiVO$_2$. It can be visualized as a rocksalt VO in which half of V ions is substituted by nonmagnetic Li. V and Li in this case order in consecutive [111] layers, and in effect we have a quasi-two-dimensional system, with V$^{3+}$ ($d^2$) ions forming triangular layers separated by similar layers of nonmagnetic Li, see Fig. ~\ref{fig:LiVO2}(a).

LiVO$_2$ is an insulating compound, and at T$_C\sim$460 K  it experiences structural phase transition, below which the magnetic susceptibility strongly decreases and LiVO$_2$ becomes practically diamagnetic, while it is paramagnetic above T$_C$\cite{Kobayashi1969}. This behavior was explained as being due to orbital ordering with concomittant structural distortion\cite{Pen1997}. Triangular lattice is usually considered as frustrated, meaning that it is not bipartite, i.e. it cannot be subdivided into two sublattices such that the nearest neighbors of one belong to the other. But in LiVO$_2$ we have two $d$-electrons per V which occupy triply-degenerate $t_{2g}$ orbitals, so that from the ``orbital'' point of view it is a triply-degenerate system. These three $t_{2g}$ orbitals are shown (by different colors) in Fig.~\ref{fig:LiVO2}. And a triangular lattice, though it cannot be subdivided into two sublattices, can be naturally subdivided into three! This is what indeed happens in LiVO$_2$ below T$_C$. The orbital ordering proposed for LiVO$_2$ in Ref.~\cite{Pen1997} is shown in Fig. ~\ref{fig:LiVO2}(b).  We see that due to this orbital ordering the system is subdivided into tightly-bound triangles (shaded in Fig.~\ref{fig:LiVO2}(b)). According to Goodenough-Kanamori-Anderson rules one would have in these trimers a strong antiferromagnetic exchange (between the half-filled $t_{2g}$ orbitals), whereas the exchange between these trimers would be very weak and presumably ferromagnetic. In any case, antiferromagnetic coupling between V ions in these triangles, each V with $S=1$ (two $d$-electrons per V), would make a spin-singlet ground state (three spins 1 form total singlet (so to say, ``1+1+1=0'').

Indeed, representing the Heisenberg Hamiltonian for a triangle as
\begin{eqnarray}
H &=& 2J \left( {\mathbf S}_1 {\mathbf S}_2 + {\mathbf S}_1 {\mathbf S}_3 +  {\mathbf S}_2 {\mathbf S}_3 \right) =J \left( {\mathbf S}_1+ {\mathbf S}_2 + {\mathbf S}_3 \right)^2\nonumber \\
&-& J \sum_{i=1}^3 {\mathbf S}_i^2 = J {\mathbf S}_{tot}^2- J \sum_{i=1}^3 {\mathbf S}_i^2,
\end{eqnarray}
we see that for the antiferromagnetic coupling $J>0$ the ground state corresponds to a total spin $S_{tot}=0$.

Similar conclusion we would get if we treat $d$-electrons in these trimers as itinerant: in this case these trimers would form just a triangular molecule with  singlet dimers at each bond of a triangle, formed by respective orbitals, Fig.~\ref{fig:LiVO2}(b), with the
$S_{tot}=0$ ground state. This picture would be more applicable if the effective $d-d$ hopping within these trimers would be larger than the Hund's rule coupling on each V, i.e. $t > J_H= 0.8-0.9$ eV. Which of these two limiting pictures is closer to reality in LiVO$_2$ is still an open question.  Spectroscopic studies seem to be in favor of the first interpretation (localized electrons forming spin $S=1$ at each V, which are coupled to total $S_{tot}=0$ in a trimer)\cite{Pen1997}. However, structural distortion accompanying this transition in LiVO$_2$ leads to the formation of very short V-V bonds in such trimers: V-V distance in these is  2.56 \AA - even shorter that the V-V distance of 2.62 \AA~in V metal \cite{james1960lattice}! From this point of view one could expect that the better description of V trimers can be obtained in a picture of electrons ``delocalized'' within each trimer. Further studies, both experimental and theoretical, could be very helpful to resolve this dilemma.

\begin{figure}[t]
   \centering
  \includegraphics[width=0.49\textwidth]{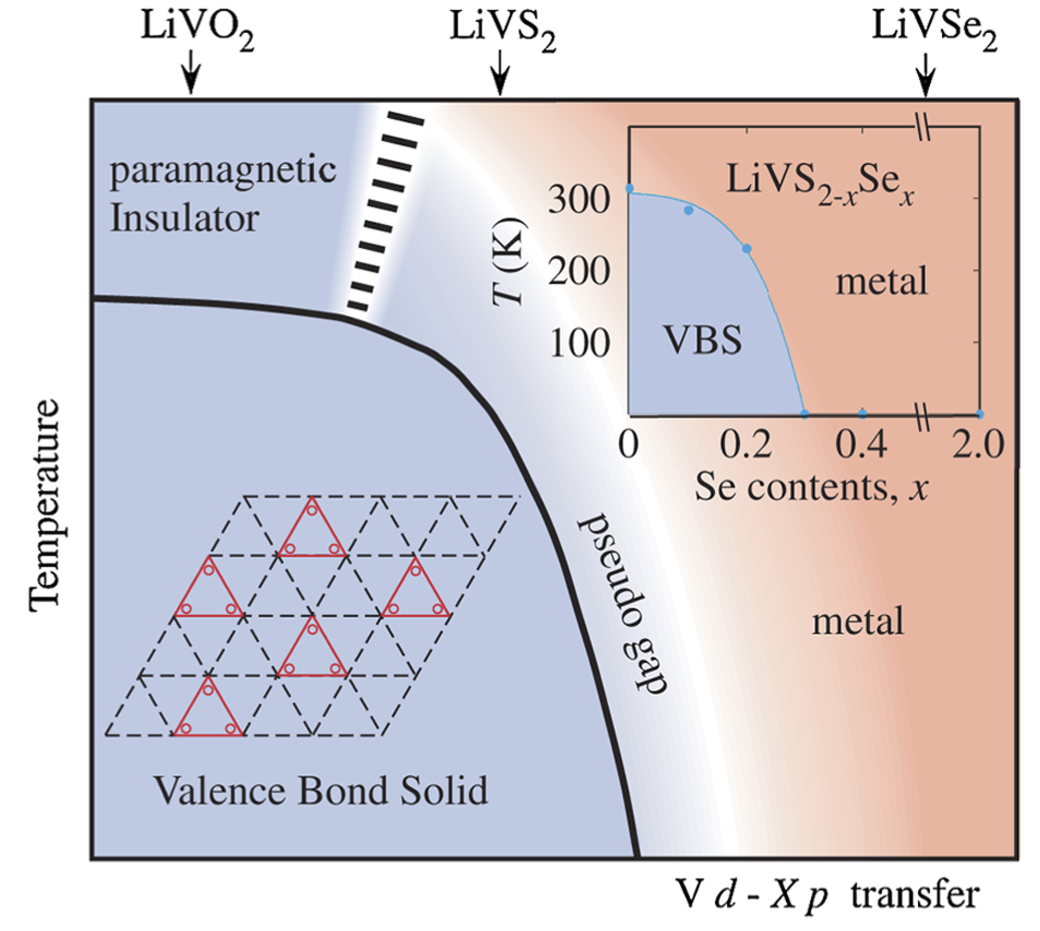}
  \caption{\label{fig:LiVO2-PD} Schematic phase diagram of LiV$L_3$, where $L$ is O, S, or Se. Taken from \cite{Katayama2009}.}
\end{figure}

An important information about the formation of clusters close to Mott transition was provided by the experiments by the group of Takagi\cite{Katayama2009}. The authors extended the study of this phenomenon, observed in  LiVO$_2$, to LiVS$_2$ and LiVSe$_2$ with the same structure, but with stronger covalency than in oxide, see Fig.~\ref{fig:LiVO2-PD}.  LiVS$_2$ has similar transition from the undistorted state to the  diamagnetic state with the same trimers as in LiVO$_2$. But in this case it is a real metal-insulator transition: LiVS$_2$ is a metal above T$_C$, but becomes insulator in the trimerized phase below T$_C$. Going further to LiVSe$_2$ one reaches real metallic state which survives down to $T=0$. Thus in these systems we have spanned the whole series: insulator-insulator transition in LiVO$_2$, metal-insulator transition in LiVS$_2$, and a homogeneous metallic state in LiVSe$_2$. Apparently the formation of these tightly bound trimers in LiVO$_2$ and LiVS$_2$ is intrinsically connected with the proximity to such localized-itinerant crossover, or to a Mott transition, and can be seen as a precursor of such transition.

The example of LiVO$_2$ also clearly shows that the ``molecules'' formed close to Mott transition can be not only dimers, which we met e.g. in VO$_2$\cite{Imada1998} or NaTiSi$_2$O$_6$\cite{Isobe2002}, but they can be larger clusters – in this case trimers V$_3$. There are also examples of still larger molecular clusters formed in this situation. For example tetramers are formed in CaV$_4$O$_9$\cite{Korotin1999}. One can also speak about tetramer molecules in the so called lacunar spinels like GaV$_4$O$_8$, which can be visualized as distorted A-site deficient spinels Ga$_{1/2}$(Vacancy)$_{1/2}$V$_2$O$_4$. In this case one can very successfully describe their electronic structure by molecular orbitals at respective clusters, and such ``molecules'' can even form Mott insulators, with  these clusters playing the role of sites in the Mott-Hubbard description of these systems\cite{Abd-Elmeguid2004,Harris1989}. Actually a very similar situation exists also in pure and doped buckyballs C$_{60}$,  for example in K$_3$C$_{60}$, where electrons ``live'' on molecular orbitals of C$_{60}$ balls, and, depending on the occupation of respective molecular levels, we may have either singlet (``low-spin'') states, or states with magnetic moments localized on such molecules\cite{Fabrizio1997}. And at certain conditions we may have here insulator-metal transitions, and in metallic state the materials can even become superconducting\cite{Palstra1995}.
\begin{figure}[t]
   \centering
  \includegraphics[width=0.49\textwidth]{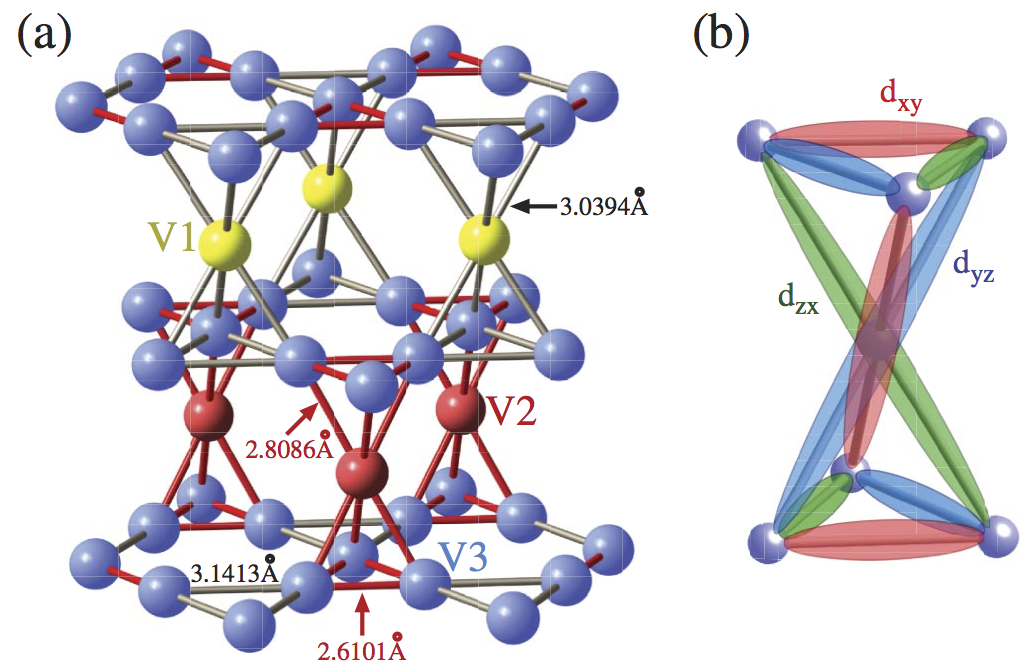}
  \caption{\label{fig:AlV2O4} (a) Crystal structure of AlV$_2$O$_4$ (only V ions are shown). Vanadium heptamers formed by short V-V bonds are shown in red. (b) Suggested molecular orbitals, which result in formation of these heptamers. Reproduced from Ref.~\cite{Horibe2006}.}
\end{figure}

There exist also other systems with similar properties. Even larger such molecular clusters are formed in a spinel AlV$_2$O$_4$, where below the metal-insulator transition there occurs a structural deformation with the formation of V heptamers – clusters comprising 7 V ions, see Fig.~\ref{fig:AlV2O4}\cite{Horibe2006}. And, similar to LiVO$_2$, at least some V-V bonds in these ``molecules'' are also shorter than those in V metal.

Sometimes one can use this concept of ``molecules'' in solids also for systems in which there are no such clusters structurally. Even in this case there can be situations, in which electronically one can describe a system as composed of ``molecules''.

\begin{figure}[t]
   \centering
  \includegraphics[width=0.42\textwidth]{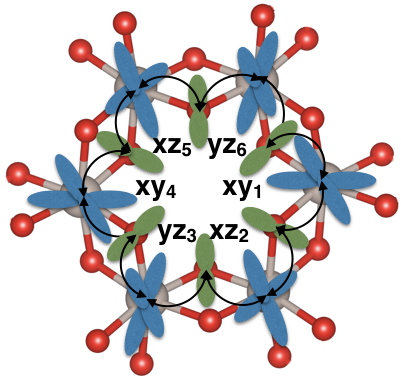}
  \caption{\label{fig:SrRu2O6} Formation of quasimolecular orbitals on honeycomb lattice via $p-d$ hopping in SrRu$_2$O$_6$. Transition metal ions are shown by grey, while ligands by red balls. Starting from one of the $t_{2g}$ orbitals (blue), the  $d$-electron turns out to be confined in the quasimolecular orbital on one of the hexagons, if only the hopping via $p$ orbitals (green) of ligand are taken into account.}
\end{figure}

The honeycomb geometry is very interesting from this point of view. We consider TM ions in octahedral coordination with not completely filled $t_{2g}$ shell. These octahedra form honeycomb lattice sharing their edges, as, e.g., in Na$_2$IrO$_3$ or SrRu$_2$O$_6$. If one includes hoppings via ligand $p$ orbitals, then due to signs of the wavefunctions in such a geometry the $d$-electron can hop only within one particular hexagon and cannot move to another one, as shown in Fig.~\ref{fig:SrRu2O6}. I.e. if we start for instance from the $xy_1$ orbital on a site 1, then the electron can hop only to the $xz_2$ and $yz_6$ orbitals of neighboring TM ions in a certain hexagon (indexes numerate TM ions). Being on these orbitals it cannot escape this TM$_6$ hexagon, but can only move to the $yz_3$ and $xz_5$ orbitals and so on. Thus, the nature of electrons in this case is twofold\cite{Streltsov2015a}. On one hand they are itinerant within hexagon, but on the other --  localized on  some extended orbitals, which were called quasimolecular orbitals (QMOs)\cite{Mazin2012}. It is interesting that QMOs give the band spectrum, which reminds the electronic spectrum of a benzene molecule. 
 
This type of description of the electronic structure of TM oxides having honeycomb lattice was first proposed for Na$_2$IrO$_3$ and Li$_2$IrO$_3$~\cite{Mazin2012,Foyevtsova2013}. However, it turned out that in iridates this model is still not perfect, there are effects which lead to ``mixing'' of these QMOs, such as the direct $d-d$ hopping and the SOC (see Sec.~\ref{sec:Kitaev} for details). But for example in SrRu$_2$O$_6$ this picture works much better~\cite{Streltsov2015a}. In SrRu$_2$O$_6$ the presence of these QMOs is expected to strongly affect optical properties\cite{Pchelkina2016} and can also be important for description of its unusual magnetic properties, in particular a very high for the layered material Neel temperature of $\sim560$K \cite{Hiley2015,Tian2015}. 

\begin{figure}[t]
   \centering
  \includegraphics[width=0.49\textwidth]{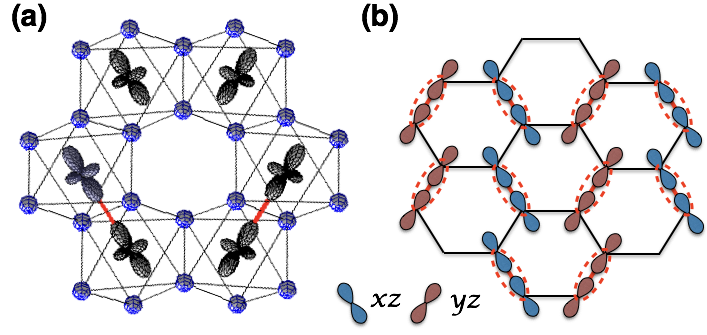}
  \caption{\label{fig:Li2RuO3} (a) Orbitals providing strong direct $d-d$ bonding and finally leading to the formation of the spin singlet state in Li$_2$RuO$_3$ (the results of the GGA calculations). (b) Herringbone distribution of these singlets, found in the low-temperature phase of Li$_2$RuO$_3$\cite{Miura2007,Miura2009}.}
\end{figure}

It is clear that the  direct $d-d$ and the $p-d$ hoppings on honeycomb lattice would stabilize very different states. The $p-d$ hopping may result in the formation of QMOs, living on hexagons (Fig.~\ref{fig:SrRu2O6}), while the direct $d-d$ hopping would favor strong metal-metal bond on particular two-site bonds (Fig.~\ref{fig:Li2RuO3}). One sees that in this case an electron put on such $t_{2g}$ orbital, can only hop to one nearest neighbor and back. Thus the effective dimensionality in this case would be reduced from 2D to 0D! This is a clear example of reduction of effective dimensionality due to orbital ordering discussed in Sec.~\ref{sec:red-dim}.
Indeed, there is one $t_{2g}$ orbital at each site in the common edge geometry,  directed towards a neighbor,  which would give such a strong bonding, cf. Fig.~\ref{fig:packing}(b). 

Relative importance of the $d-d$ or $p-d$ hoppings depends on a particular situation. In the case of large metal-metal distance the main hopping would occur via ligands (since $t_{dd}$ falls drastically with distance \eqref{eq:dd-r-dependence}); but for short distances the direct $d-d$ hopping may dominate. It seems that, for example, Na$_2$IrO$_3$ belongs to the first class of systems, whereas Li$_2$IrO$_3$, with smaller Li ions, may already be ``half-way'' to the second case. This, in particular, may be responsible for a more complicated magnetic structure of Li$_2$IrO$_3$ as compared with Na$_2$IrO$_3$\cite{Biffin2014}. Distortions of the octahedra and the SOC can also intervene.

Speaking of honeycomb systems, it is interesting to compare the situation in ``213'' iridates like Na$_2$IrO$_3$, and a similar system with Ru instead of Ir, Li$_2$RuO$_3$.  Li$_2$RuO$_3$ may be an example of a system of the second type, in which the direct $d-d$ hopping may be more important than that via oxygens. Whereas Na$_2$IrO$_3$ and Li$_2$IrO$_3$ remain undistorted and at low temperatures they develop long-range magnetic ordering\cite{Biffin2014},  in Li$_2$RuO$_3$ there occurs below T$_C \sim 540$ K  a phase transition with the formation of diamagnetic Ru dimers\cite{Miura2007,Miura2009}, with  the Ru-Ru distance in a dimer being rather short, 2.57 \AA\cite{Miura2007} -- again shorter that that in Ru metal (2.65 \AA). These dimers form below T$_C$ an interesting herring-bone pattern. The formation  of such dimers, as explained in Ref.~\cite{Jackeli2008}, is a consequence of an orbital ordering, with the direct $d-d$ hopping playing the main role, see Fig.~\ref{fig:Li2RuO3}.  Ab-initio calculations in general support this picture, although it seems that in reality also the hopping via oxygens is not negligible. Interestingly enough, the dimer Ru-Ru ``molecules'',  ordered in Li$_2$RuO$_3$ below 540 K, are very stable, and they persist even above T$_C$, in the ``average'' hexagonal phase. Pair distribution function (PDF) study has demonstrated that they survive locally up to at least 650 C, forming disordered, and probably dynamic (liquid-like) state – a dimer liquid\cite{Kimber2013}. The NMR data also detect thermal activation processes associated with the flow of dimers\cite{Arapova2017}. The study of this system for different stoichiometry supports this conclusion\cite{Park2016}.

The general question, in which cases in such honeycomb systems one ends up in an undistorted magnetic state, and when it is more favorable to form singlet dimers ordered in a particular pattern, is an interesting and still an open question. As we just argued, one can give qualitative arguments that when the direct $d-d$ hopping  dominates, one can have better conditions for the formation of “molecular” state (another name for such state is a valence bond solid \cite{sachdev}). The dominant hopping via oxygen $p$ orbitals may work rather in favor of less localized states, although the notion of molecular, or rather quasimolecular orbitals may be applicable in such cases too.

 In any case, all these examples demonstrate that indeed there may appear novel states close to localized-itinerant crossover, so that the Mott transition occurs ``step-wise'': first the electrons are delocalized in finite clusters, forming ``molecules'' in a solid – the hopping between such molecules being still small enough to render the whole system insulating, but with electrons localized rather on such “molecular clusters” and not on isolated sites. And only later, for example at still much higher pressures, can one reach a state of a homogeneous metal, in which electrons would be really itinerant, delocalized over the whole system. This is of course not a universal behavior – for example it strongly depend on the lattice geometry  (being less plausible for systems like perovskites with corner-sharing MO$_6$ octahedra); but in many cases one indeed should expect, and really observes such behavior.

\section{Orbital-selective effects\label{sec:OS-large}}
\subsection{Orbital-selective Mott transition\label{sec:OSMT}}
\begin{figure}[tbp]
   \centering
  \includegraphics[width=0.45\textwidth]{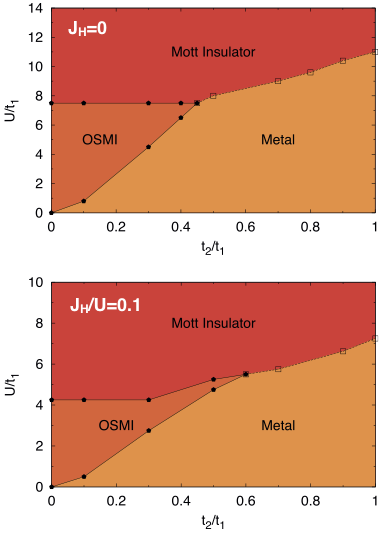}
\caption{\label{OSMT-PD} Phase diagram of the two-band nondegenerate Hubbard model on square lattice illustrating the onset of the orbital-selective Mott insulating (OSMI) phase. Top panel corresponds to $J_H=0$, while the bottom one -- to $J_H/U=0.1$. Reproduced from Ref.~\cite{Tocchio2016}.}
\end{figure}
Generally speaking, a separation of all electrons on those which behave like itinerant, and others, which are more localized, can occur not only in real space due to formation of finite size clusters, but equally well a system may stay uniform even on a small scale, but have electrons of a very different character: ``insulating'' and ``metallic''.  In other words, the Mott transition can occur not simultaneously for all bands, but in turns, i.e. it can be orbital-selective. The term ``orbital-selective Mott (OSM) transition'' was coined in Ref.~\cite{Anisimov2002a} for the description of electronic properties of Ca$_{2-x}$Sr$_x$RuO$_4$, when it was found that the transition to an insulating state for the narrow $xz/yz$ bands occurs at much smaller values of $U$ than for the $xy$ band having larger bandwidth. Thus in the regime of large $U$ the whole system is insulating due to correlation effects, for small $U$ it is metallic, but in the intermediate regime some of the electrons are localized, while the others are itinerant.

Since there are two very different species of electrons, one needs to use  the Hubbard model with inequivalent bands to describe such a situation.  The simplest would be the two-band Hubbard model with different nearest neighbor hoppings $t_m$
\begin{eqnarray}
\label{OSM-ham}
H &=& -\sum_{\langle ij \rangle m \sigma} t_{m} c^{\dagger} _{im\sigma} c_{jm\sigma} +  U \sum_{im} n_{im\uparrow} n_{im\downarrow} \\
&+& U' \sum_{\substack{i,m\ne m' \\ \sigma \sigma'} } n_{im\sigma} n_{im'\sigma'} - 
J_H \sum_{\substack{i, m\ne m' \\\sigma\sigma'}}
c^{\dagger}_{im\sigma} c_{im\sigma'} c^{\dagger}_{im'\sigma'} c_{im'\sigma},     \nonumber                                                                         
\end{eqnarray}
where $i,j$ are the site and $m,m'$ are the orbital indexes, $U$ and $U'=U-2J_H$ are intra- and inter-orbital interactions, and $J_H$ is the Hund's rule exchange (one could also have different on-site energies of these two $d$-levels, e.g. due to the effect of crystal field, see below), in the 3rd and 4th terms summation runs ones over each pair of $m,m'$. The phase diagram of such a model in the case of $J_H=0$, at half-filling (i.e. for two electrons per site) and on 2D square lattice is shown in Fig.~\ref{OSMT-PD}, top panel~\cite{Tocchio2016}.   There are three main regions: (1) homogeneous metallic state when $t_2/t_1 \to 1$ and for small $U$; (2) insulating Mott phase for large $U$; and (3) an intermediate OSM phase (one ignores here possible complications like the eventual formation of the spin density wave (SDW) state due to nesting of the Fermi-surface, which could  appear even for small $U$).

It is important to mention that the phase diagram presented in Fig.~\ref{OSMT-PD} was obtained for an ideal situation, when there is no mixing between two orbitals in the kinetic energy term. Any hybridization between these orbitals, i.e. the presence of the terms like  $t_{mm',ij} c^{\dagger} _{im\sigma} c_{jm\sigma}$ with $m \ne m'$ would suppress OSM state. Such terms are always present in real systems and they will disfavor OSM phase. However, there are also other factors which, in contrast, stabilize such a state. First of all, the Hund's rule intra-atomic exchange $J_H$ suppresses any orbital fluctuations irrespective of $U$ and thus supports OSM state, as one can easily see comparing the top and bottom panels of Fig.~\ref{OSMT-PD}. There are also other factors, which help to decouple different orbitals. For example, it was shown in Ref.~\cite{Medici2009} that the OSM phase may appear even in the situation when two bands have the same bandwidths (i.e. $t_1=t_2$), but there is a crystal-field splitting supported by the Hund's exchange.

In the same way as the formation of ``molecules'' in homogeneous solids (described in Sec.~\ref{sec:molecules}), the OSM phase is a precursor of a phase transition. It is important to mention that the OSM phase is not simply a theoretical toy, but that the orbital selectivity strongly affects physical properties. Thus the orbital-selective localization leads to a non-Fermi-liquid behaviour\cite{Biermann2005a}. Moreover, it is well known that an insulating state can be obtained only for integer site occupancies in the Hubbard model, and any doping makes a system metallic. In contrast, the OSM phase is robust against doping\cite{Koga2004}. This can be easily rationalized, since doping changes only a position of the chemical potential $\mu$ within the metallic band formed by itinerant electrons, and the OSM phase is stable until the total change of $\mu$ exceed the energy gap provided by localized electrons.

Very recently the influence of the electron-phonon interaction on the OSM state was studied in the frameworks of the Hubbard-Holstein model with two electronic bands having very different bandwidths ($t_1/t_2=5$)\cite{Li2017}. In particular it was found that if we change the strength of the electron-phonon interaction, then the transition from the uniform metallic state to the phase with the charge-density wave (CDW) also occurs through the orbital-selective phase (with site-centered CDW).

It has to be mentioned that the idea of OSM state was implicitly used long before Ref.~\cite{Anisimov2002a}. Indeed, for example in order to explain double exchange mechanism of the ferromagnetism one needs to treat part of electrons as itinerant, moving on the background of localized magnetic moments provided by completely different electrons, which essentially do not hop from site to site (see Sec.~\ref{Sec:DE}). This is actually the picture always used to describe for example the properties of the colossal magnetoresistance manganites La$_{1-x}$Sr$_x$MnO$_3$, La$_{1-x}$Ca$_x$MnO$_3$\cite{Tokura1999}. For these systems one usually treats electrons on the half-filled $t_{2g}$ shell ($t_{2g}^3$) as localized, and the electrons in the $e_g$ bands as itinerant. This picture was already described in Sec.~\ref{Sec:DE}. One could in principle include correlation effects also for the $e_g$ electrons, but even without  those, the purely itinerant picture of the $e_g$ electrons gives a very reasonable description of many properties of these manganites\cite{Efremov2005}. The same description ($t_{2g}$ electrons localized, $e_g$ itinerant) can be also successfully used for other systems with perovskite and perovskite-related structures. The ideas similar to the OSM state were used in the Kondo physics, e.g. for the description of the heavy-fermion compounds, for which usually the electrons of different shells are considered as localized (typically $f$) and mobile ($s,p,d$).

\subsection{Orbital-selective behaviour and (partial) suppression of magnetism\label{sec:OS}}
\begin{figure}[tbp]
   \centering
  \includegraphics[width=0.49\textwidth]{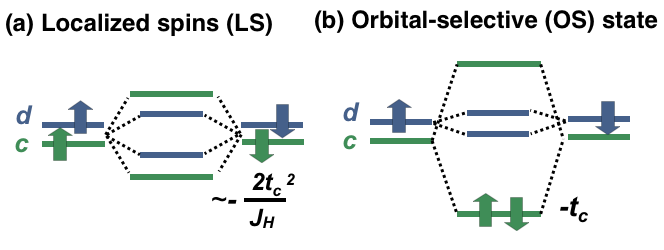}
  \caption{\label{OS-4e} Energy scheme illustrating formation of the orbital-selective state in case of two orbitals and two electrons per site (b). In the left panel, (a), the conventional state with spin 1 per site (due to dominating Hund's intra-atomic exchange) is presented.}
\end{figure}

We have already seen that the formation of molecular orbitals, promoted by corresponding orbital occupation,  can weaken and even completely suppress magnetism  in some materials - e.g. in NaTiSi$_2$O$_6$, CuIr$_2$S$_4$, and LiVO$_2$ (see Sec.~\ref{sec:molecules}).
But one can also anticipate the situation in which the electrons on one orbital form singlet state, whereas other electrons still remain localized and contribute to magnetism - albeit with strongly reduced moment. Or the other electrons can be delocalized, but not forming singlet dimers.  Such situation would be in some sense analogous to the orbital-selective behavior described in previous subsection. One can illustrate it on a simple model, which,  as we show below, actually rather closely corresponds to the experimental situation in some real materials.

Consider a dimer with two orbitals on each site, with strong intersite hopping $t_c$ of one orbital, call it $c$, and no (or very small) hopping of the other, $d$-orbital, $t_d$, 
\begin{eqnarray}
H = -\sum_{\langle ij \rangle \sigma} (t_c c^{\dagger}_{i \sigma}c_{j \sigma} + t_d
d^{\dagger}_{i \sigma}d_{j \sigma}) -J_H \sum_i (\frac 12 + 2 S^z_{id} S^z_{ic}). \nonumber
\end{eqnarray}
and consider the case of two electrons per site. If the Hund's coupling $J_H$ is the largest parameter in the system, first both electrons at each site form, according to the first Hund's rule, the state with the spin $S=1$, see Fig.~\ref{OS-4e}(a). Then these electrons would have some exchange because of virtual hopping of $c$ electrons between sites, which would give antiferromagnetic coupling of these spin 1 sites with the exchange constant   $J \sim 2t_c^2/J_H$  (cf. the usual expression \eqref{t2U} for the  exchange interaction in simple Hubbard model; here we have not yet included the Hubbard repulsion $U$, but the virtual state with an electron transferred to a neighboring site has an excitation energy $J_H$, which now stands in the denominator of the expression for $J$ instead of the Hubbard's $U$ in \eqref{t2U}). If $t_d=0$, the energy of this state is thus 
\begin{eqnarray}
\label{eq:1}
E_{LS} = -2J_H - 2t_c^2/J_H.           
\end{eqnarray}
But for smaller $J_H$ and large enough hopping $t_c$ we can have a very different state, Fig.~\ref{OS-4e}(b): We can make a singlet from the $c$ electrons, breaking the $S=1$ states at each site, stabilized by the Hund's interaction. We thus lose (large part) of the Hund's energy. But instead of that now these $c$ electrons can gain bonding energy $-2t_c$. The energy of such state would be
\begin{eqnarray}	
\label{eq:2}
E_{OS} = -2t_c - J_H            
\end{eqnarray}
for $t_d=0$ (part of the Hund's energy we still gain when $c$ and $d$-electrons are at the same site with their spins parallel). In any case, comparison of these expressions \eqref{eq:1} and \eqref{eq:2} shows that this second state, with two electrons occupying molecular-orbitals formed by the $c$ orbitals, is more favorable if 
\begin{eqnarray}	
t_c > J_H/2.                     
\end{eqnarray}
In this state the  remaining $d$-electrons, one per site, would live their own life irrespective of the $c$ electrons, for example they can make magnetic ordering, but with strongly (here twice) reduced magnetic moment: spin 1/2 per site instead of spin 1 in case of dominating Hund's coupling. Therefore, this state can be called orbital-selective\cite{Castellani1978}.

We have mentioned that the idea of orbital selectivity lies at the heart of the double exchange, but how this differentiation on the $c$ and $d$-orbitals may occur? In fact, this is a very natural situation in many geometries. For example, the $xy$ orbitals will have much larger direct hoppings than $xz$ or $yz$ in a common edge geometry, see Fig.~\ref{fig:packing}b, or the $a_{1g}$ orbitals overlap much stronger than $e_g^{\pi}$ in a common face case, see Fig.~\ref{fig:packing}c. This is the reason why orbital-selectivity is not such a  rare phenomenon. Let's consider for example  $\alpha-$MoCl$_4$, where Mo$^{4+}$ ions have $4d^2$ configuration. One might expect that the effective Curie-Weiss magnetic moment in this situation would be $\mu_{eff} \sim 2.8 \mu_B$, but in fact it turns out to be much smaller, $\mu_{eff} \sim0.9\mu_B$\cite{Larson1964,Kepert1968}. This is because of the large overlap between the $xy$ orbitals (=$c$ orbitals), which form singlet molecular orbitals,  so that the magnetic moment is provided only by the electrons occupying $xz/yz$ orbitals (=$d$ orbitals).\cite{Korotin2016}. 
\begin{figure}[t]
   \centering
  \includegraphics[width=0.49\textwidth]{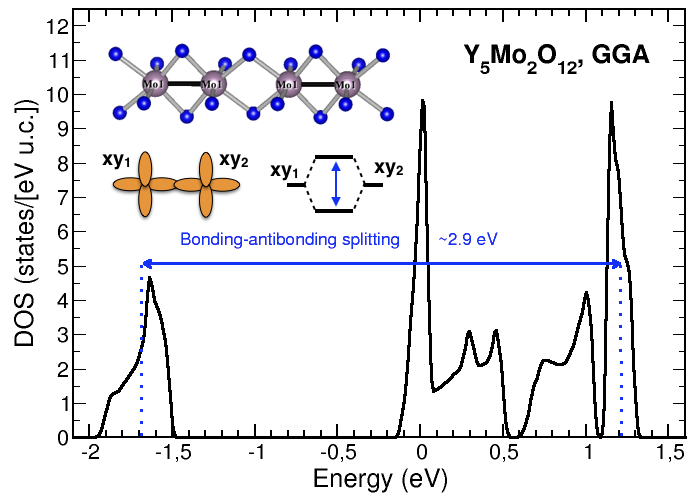}
  \caption{\label{YMoO} The density of state as obtained in the generalized gradient approximation for Y$_2$Mo$_5$O$_{12}$\cite{Streltsov2015MISM}. Crystal structure and $4d$ orbitals of Mo, which strongly overlap, are shown in the insets.}
\end{figure}

Another example of this behaviour is provided by rutile systems VO$_2$ and MoO$_2$. In VO$_2$ (V$^{4+}$, $d^1$) the famous metal-insulator transition at 68 $^{\circ}$C is accompanied (or is driven by) the formation of V-V dimers in chains in the $c$ direction, where VO$_6$ octahedra share common edge, the dimers being formed by corresponding $xy$ orbitals (in local coordination system). In contrast, MoO$_2$ (Mo$^{4+}$, $d^2$) remains metallic down to $T=0$. Nevertheless structurally MoO$_2$ develops the same Mo dimers as V dimers below T$_C$ in VO$_2$! Apparently, the electrons on the $xy$ orbitals in MoO$_2$ form such dimers, whereas the other electron per Mo behaves quite differently, in this case forming a metallic band. This is also a very clear example of then orbital-selective behaviour.

\begin{figure}[tbp]
   \centering
  \includegraphics[width=0.49\textwidth]{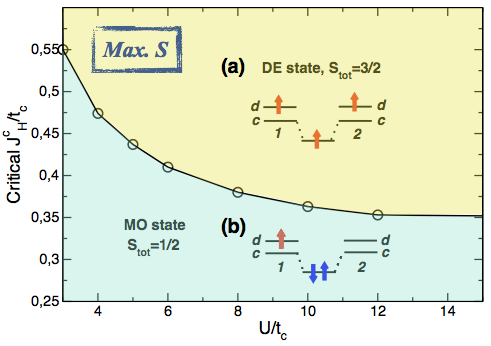}
  \caption{\label{DE:FD} Phase diagram for a dimer with 2 orbitals and 1.5 electrons per site in $U$ (on-site Hubbard repulsion) and $J_H$ (Hund's rule coupling) coordinates. Results of the exact diagonalization at $T=0$ K\cite{Streltsov2016b}.}
\end{figure}

It is important to mention that the orbital-selective behaviour can be seen not only in the case of integer number of electrons per site, but also for other fillings. For example in the case of an isolated dimer with 3 electrons (1.5 electrons per site) one can easily find the energies of the two energetically lowest solutions. The first one, shown in the inset (a) of Fig.~\ref{DE:FD}, is a ``molecular'' version of the DE: the $c$ electron hops from site to site and forces $d$ electrons to have the same spin projection (we chose $t_d=0$ for simplicity):
\begin{eqnarray}
E_{DE} = -J_H - t_c.
\end{eqnarray}
This state has the maximal total spin $S_{tot}=3/2$ (it corresponds to the ferromagnetic order in the conventional DE).

However, there is also a different state with $S_{tot}=1/2$, sketched in the inset (b) of Fig.~\ref{DE:FD}. In this state two electrons are in the bonding state constructed out of the $c$-orbitals. This state is stabilized by a large hopping between  $c$-orbitals, $t_c$. It is an orbital-selective (OS) state in a sense that only part of the orbitals ($d$) provide spin moment, while electrons on other orbitals ($c$) form a singlet state. The energy of this state is
\begin{eqnarray}
E_{OS} = -\frac{J_H}2 - 2t_c.
\end{eqnarray}

We see that these two states will compete, and the total spin of the system can be suppressed if 
\begin{eqnarray}
\label{OS-DE-competition}
2t_c > J_H.
\end{eqnarray}
While in the double exchange-like treatments $J_H$ is typically treated as the  leading parameter, and the condition \eqref{OS-DE-competition} can be considered as unrealistic, in real materials it can be easily fulfilled.  As was already mentioned above, this may be the case in some $4d$ and $5d$ systems, for which the Hund's coupling is weaker, but the extension of $d$ functions and with it the value of intersite hopping $t$ increases.

One of such examples is Y$_5$Mo$_2$O$_{12}$, which has the structure of dimerized chains\cite{Torardi1985}, 
and in which Mo ion has the  $4d^{1.5}$ electronic configuration, the same as considered above. The dimers are formed by the edge-sharing MoO$_6$ octahedra, see inset in Fig.~\ref{YMoO}. There is a very strong overlap between the $xy$ orbitals in this geometry. Corresponding bonding-antibonding splitting exceeds 2.9 eV, and $t_{xy/xy}\sim 1.4$ eV, while $t_{xz,yz/xz,yz}\sim 0.3$ eV\cite{Streltsov2015MISM}.  Thus, we see that $2t_c \approx 2.8$ eV is much larger than any possible values of the Hund's coupling $J_H$ (typically $\sim 0.5-0.7$ eV for the $4d$ elements).   Thus in this case, as in our toy model, the $xy$ orbitals form singlet state on a dimer, which  results in a considerable reduction of the magnetic moment observed in this system: $\mu_{eff}^{exp} = 1.7 \mu_B$/Mo\cite{Torardi1985}, which is much smaller than $\mu_{eff}^{theor} = 2.3 \mu_B$/Mo expected for Mo$^{4.5+}$. 

The situation very similar to that in Y$_5$Mo$_2$O$_{12}$ is also observed in Y$_5$Re$_2$O$_{12}$, with the same crystal structure\cite{Chi2003}. In this system, with the Re valence 4.5+ (electronic configurations $d^2/d^3$) the moment per dimer is again strongly reduced, even stronger than in Y$_5$Mo$_2$O$_{12}$: it correspond to $S=1/2$ per dimer, instead of the spin 5/2 expected if the ``DE'' state would be realized. I.e. here {\it two electrons} per Re form singlet metal-metal bonds, and only one electron per dimer remains magnetic. 
\begin{figure}[t]
   \centering
  \includegraphics[width=0.49\textwidth]{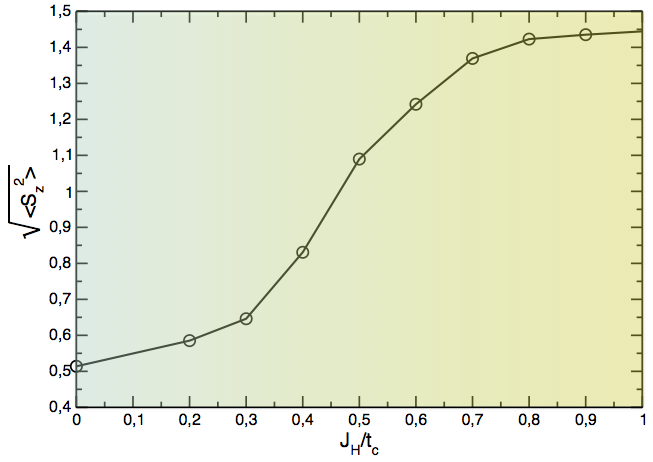}
  \caption{\label{DE:S2} Results of the cluster DMFT calculations for a dimerized chain with 2 orbitals and 1.5 electrons per site (for details see \cite{Streltsov2016b}). One may see that there is a wide crossover region, where the total spin can have any value between $S_{tot}=1/2$, corresponding to the OS state, and $S_{tot}=3/2$ of the DE solution.}
\end{figure}

The treatment of the orbital-selective formation of ``molecules’’, which was presented above, is  rather  qualitative, since it does not take into account the on-site Hubbard repulsion. But one can easily generalize it by exact treatment of a dimer case. The results of the exact diagonalization  for an isolated dimer, described by the Hamiltonian
\begin{eqnarray}
H &=& -\sum_{\langle ij \rangle \sigma} (t_c c^{\dagger}_{i \sigma}c_{j \sigma} + t_d
d^{\dagger}_{i \sigma}d_{j \sigma}) + U \sum_{i m} n_{im\uparrow} n_{i m \downarrow} \nonumber \\
&+& U' \sum_{i m \ne m'} n_{i m\uparrow} n_{i m'\downarrow} +
\frac {U'-J_H}2 \sum_{\substack{i \sigma \\ m\ne m'} } n_{i m\sigma} n_{i m'\sigma'},
\nonumber
\end{eqnarray}
in which we  also included this interaction, are presented in Fig.~\ref{DE:FD}. They clearly show that the critical ratio $J_H/t$ for the suppression of the DE solution depends on Hubbard $U$ (in the so-called Kanamori parameterization $U'=U-2J_H$\cite{Kanamori1963}).

As we have seen on the example of Y$_5$Mo$_2$O$_{12}$, the suppression of the magnetic moments due to the orbital selectivity occurs not only for isolated clusters, but it was shown to persist in dimerized systems, which can be considered as an intermediate step between isolated clusters and uniform solids. However, while in a dimer there is a discontinuous transition from the DE to the OS state (since they correspond to different quantum numbers), it becomes a smooth crossover in dimerized bulk systems, and the final value of the measured magnetization depends on specific parameters of a system under consideration, see Fig.~\ref{DE:S2}. 

It is very interesting to mention systems with the general formula Ba$_3M$Ru$_2$O$_9$, where $M$ can be In, Y, La, Lu, Nd etc (in principle one can have at these positions also the other ions, like Na$^{1+}$; Ca$^{2+}$, Co$^{2+}$; Ce$^{4+}$, Ti$^{4+}$).  Ru ions are in the RuO$_6$ octahedra, which form dimers ordered in the triangular lattice. Since Ru is 4.5+, one may expect that the local magnetic moment on Ru ion would be $\sim$2.5$\mu_B$. However, while the crystal structure of systems with different $M$-ions is almost the same\cite{Doi2001,Senn2013a}, their magnetic properties are very different\cite{Rijssenbeek1998,Doi2002}, and none of them reminds a system with the local magnetic moment of $\sim$2.5$\mu_B$. E.g. in Ba$_3$YRu$_2$O$_9$ local magnetic moment on Ru is $\sim0.5\mu_B$, while in Ba$_3$LaRu$_2$O$_9$ it is $\sim1.4\mu_B$\cite{Senn2013a}. Different models, like charge ordering (i.e. segregation on Ru$^{4+}$ and Ru$^{5+}$ ions)\cite{Doi2002} and double exchange\cite{Senn2013a} were used to explain magnetic properties of Ba$_3M$Ru$_2$O$_9$ series. In fact they can be explained using orbital-selective behaviour with the crystal structure playing a role of a fine tuner, which regulates splittings between different molecular or localized orbitals and because of this influences value of the observed magnetic moment\cite{Ziat2017}.
\begin{figure}[t]
   \centering
  \includegraphics[angle=270,width=0.53\textwidth]{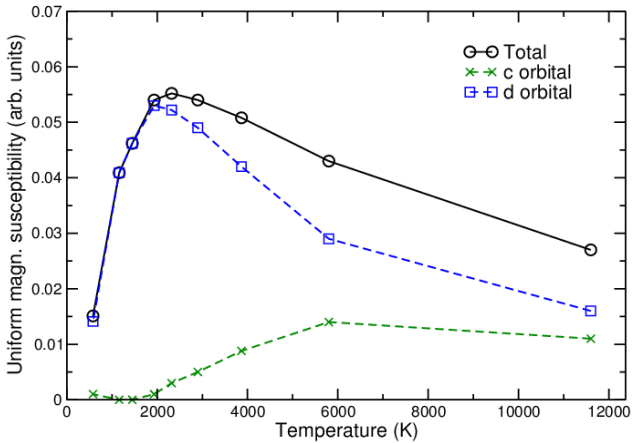}
  \caption{\label{OS:Sz} Magnetic susceptibility as obtained in the cluster DMFT calculations for a dimerized chain with two orbitals ($c$ and $d$, $t_c \gg t_d$) and two electrons per site (for details see \cite{Streltsov2014}).}
\end{figure}

\begin{figure}[t]
   \centering
  \includegraphics[angle=270,width=0.53\textwidth]{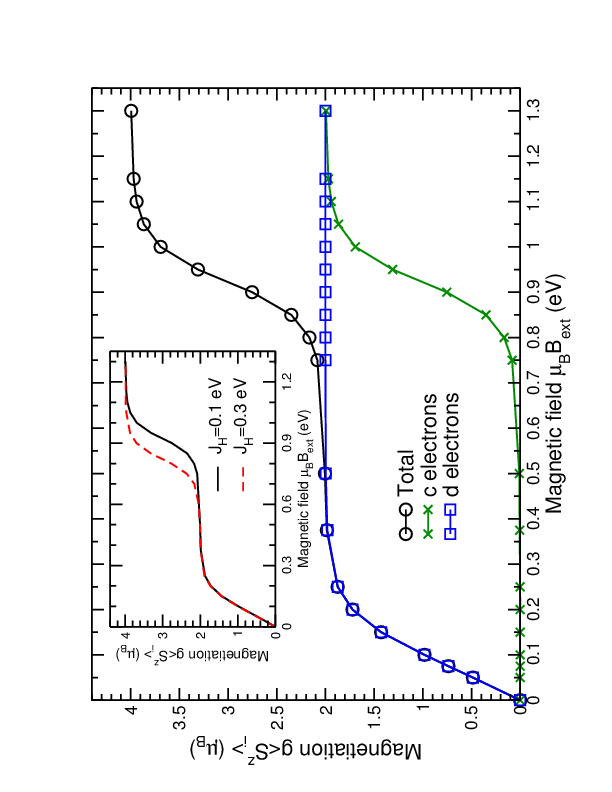}
  \caption{\label{OS:Magn} Results of the cluster DMFT calculations for a chain consisting of dimers. Magnetization dependence on applied field in the orbital-selective regime. One may see that first the localized $d$ electrons respond to the magnetic field, and only later the molecular-like $c$ electrons joint them (for details see \cite{Streltsov2014}).}
\end{figure}

One may expect many different manifestations of the orbital selectivity. For example, since there are two types of orbitals, one ($c$) having a tendency to form molecular orbitals, and another one ($d$),  electrons on which behave more like localized, these orbitals must respond very differently on external perturbations. In Fig.~\ref{OS:Sz} the temperature dependence of the magnetic susceptibility obtained in the cluster DMFT calculations for the dimerized chain with two orbitals ($c$ and $d$, again $t_c \gg t_d$) and two electrons per site is presented. One may see that the low-temperature response (which in particular determines the value of the spin gap) is due to localized ($d$) electrons only,  molecular-like $c$ electrons being in a singlet state, so that they enter the game at much higher temperature. Such behaviour may also give the plateau in the field dependence of the magnetization as one may see from Fig.~\ref{OS:Magn}\cite{Streltsov2014}.

The situation in real materials is, however, more complicated. Thus one might expect the orbital-selective behaviour in Li$_2$RuO$_3$, which was already discussed in Sec.~\ref{sec:molecules}. Because of the common edge geometry there are the strongly overlapping $xy$ orbitals (Fig.~\ref{fig:packing}b), which play a role of $c$ orbitals,  and the $xz$, $yz$ orbitals, such that $t_{xy/xy} \gg \{t_{yz,yz},t_{xz,xz}\}$. However, the LDA+DMFT calculation for this materials shows only a moderate difference between contributions of different orbital  to the magnetic susceptibility\cite{Arapova2017}. This is a result of sufficiently low symmetry (substantial distortions of the crystal structure), which leads to an orbital mixing and to a partial ``magnetization'' of the $xy$ orbitals.

While investigation of such subtle effects as different orbital contributions to the total magnetic susceptibility  requires further theoretical and experimental study, a general concept of the orbital-selectivity works very well in many dimerized systems. In Tab.~\ref{Tab:OS} one may find a number of examples for which one sees that the  theoretical magnetic moments expected from the ionic configuration of a transition metal ion are much larger than the experimental values. An agreement between theory and experiment can be achieved taking into account orbital-selective formation of molecular orbitals, which substantially reduces theoretical magnetic moments. 

One may notice that most of the TM ions in the table~\ref{Tab:OS}  are $4d$ and $5d$ TM ions, not $3d$.  The main reason for this was already discussed above:  a much larger spatial extension of the $4d$ and $5d$ wave functions as compared to  $3d$ (due to a larger principal quantum number, see e.g. \cite{Goodenough}). On  one hand, this results in the increase of the hopping parameters,  and, on the other hand, it  leads to a decrease of the Hund's exchange. Intradimer hoppings for the $4d$ and $5d$ systems may be quite large, and the resulting bonding-antibonding splitting may exceed 3 eV\cite{Streltsov2015MISM}. On the other hand,  while  typical $J_H$ for $3d$ TM are $\sim$1 eV, they are of order of $0.5-0.7$ eV for $4d$ and $0.5$ eV for $5d$ TM ions\cite{khomskii2014transition}. Both tendencies work hand in hand in stabilizing orbital-selective states.

\begin{table*}
\centering \caption{\label{Tab:OS} Examples of dimerized systems, in which the orbital-selective formation of molecular orbitals results in a considerable reduction of magnetic moments comparing to what one would expect on the basis on an ionic configuration of a transition metal ion (column: Theoretical).} 
\vspace{0.2cm}
\begin{tabular}{l|c|c|ccc}
\hline
\hline
System & Ionic & \multicolumn{2}{c}{ Local ($\mu$) or effective ($\mu_{eff}$) moment} \\
                    & conf. &  Theoretical & Experimental\\
\hline
Y$_5$Mo$_2$O$_{12}$   & $4d^{1.5}$    & $\mu_{eff} = 2.3 \mu_{B}$/Mo & $\mu_{eff} = 1.7 \mu_{B}$/Mo \cite{Torardi1985}\\
Nb$_2$O$_2$F$_3$         & $4d^{1.5}$    & $\mu_{eff} = 3.9 \mu_{B}$/dimer & $\mu_{eff}\approx 2\mu_{B}$/dimer\cite{Tran2015}\\
$\alpha-$MoCl$_4$          & $4d^{2}$    & $\mu_{eff} = 2.8 \mu_{B}$/Mo & $\mu_{eff}\approx 0.9\mu_{B}$/Mo\cite{Kepert1968}\\
Ba$_3$YRu$_2$O$_9$    & $4d^{3.5}$    & $\mu = 2.5 \mu_{B}$/Ru & $\mu = 0.5 \mu_{B}$/Ru \cite{Senn2013a}\\
Ba$_3$LaRu$_2$O$_9$   & $4d^{3.5}$   & $\mu = 2.5 \mu_{B}$/Ru & $\mu = 1.4 \mu_{B}$/Ru \cite{Senn2013a}\\
Ba$_5$AlIr$_2$O$_{11}$ & $4d^{4.5}$    & $\mu_{eff} = 3.3 \mu_{B}$/dimer & $\mu_{eff} \approx 1 \mu_{B}$/dimer\cite{Terzic2015,Streltsov2017}\\
\hline
\hline
\end{tabular}
\end{table*}

\section{Spin-orbit related effects \label{sec:SOC}}

\subsection{Spin-orbit coupling vs. Jahn-Teller effect}
As we saw above, strong effects due to the SOC are expected for partially-filled $t_{2g}$ bands, for which orbital moment is not quenched. In this case one generically has a (triple) orbital degeneracy,  and the concomitant Jahn-Teller effect - structural distortion with the decrease of symmetry, should lift this orbital degeneracy. But we  saw that the SOC also chose particular orbital occupation, which can also lift orbital degeneracy. An important question is what is the possible interplay of the JT effect and the SOC in different situations. 

In many cases, for systems with large spins, one can get a reasonable description of the role of the SOC and the interplay between the SOC and the JT-driven orbital ordering, if instead of \eqref{eq:full-SOC} one treats the SOC classically, or in the mean-field approximation, keeping only the term $\lambda l^z  s^z$. This can be done for several ions from the end of the $3d$ row for which typically we have high spin state. One can have in such cases the situation with partially-filled $t_{2g}$ bands and simultaneously large total spin. Such is, for example, the situation in Co$^{2+}$  ($t_{2g}^5 e_g^2$, $S=3/2$) or Fe$^{2+}$ ($t_{2g}^4 e_g^2$, $S=2$). One can easily show that in this case for partially-filled $t_{2g}$ levels the SOC and  the JT distortions lead to the opposite splitting of the $d$-levels and to the opposite distortions of MO$_6$ octahedra. 

\subsubsection*{Weak SOC, mean field treatment}
Consider for example the case of one ``extra'' $t_{2g}$ electron with the spin down (while there are also 5 electrons in the spin-up channel giving large total spin) and octahedral geometry,  like e.g. for the high-spin state of Fe$^{2+}$. The JT effect would lift three-fold orbital degeneracy in such a way that the doubly-occupied $xy$ level goes down  (by $E_{JT}$),  and the half-filled doubly-degenerate $xz$ and $yz$ levels go up (by $E_{JT}/2$). This will correspond to  a tetragonal compression of MO$_6$ octahedra, Fig.~\ref{fig:JT-elong}(a).  We gain by that the JT energy $-E_{JT}$. But the occupied $xy$ orbital by that is the orbital with $l^z_{eff} = 0$, i.e.  the SOC coupling $\lambda l^z s^z$ does not give any energy gain in this case.

If instead we have not local compression, but  local elongation of MO$_6$ octahedron, the level structure is that shown in Fig.~\ref{fig:JT-elong}(b). Doubly-degenerate $xz$ and $yz$ levels, or their complex linear combinations $|l^z = \pm 1 \rangle  = \frac 1 {\sqrt 2}(xz \pm i yz)$ go down by the energy -$E_{JT}/2$. But now the SOC can lead to further splitting of these levels, by the value $\lambda$, i.e. the ground state energy gain for the ``extra'', sixths electron in this case is $E = -\frac 12 E_{JT} – \frac 12 \lambda$. Thus in this case the deformation would go by the ``JT route" (local contraction, $c/a<1$) if $E_{JT} > \lambda$, and by the ``SOC route'' (local elongation, $c/a>1$) in the opposite case. We see that in this case the JT effect and the SOC tend to cause distortions of opposite type, and they lead to occupation of different orbitals. One can show that the same situation also exists for other fillings of $t_{2g}$ levels and even in other local surroundings, e.g. in tetrahedra, see~\cite{Goodenough}.

Experiment shows that for the  heavier $3d$ elements, such as Fe, Co etc, usually ty larger SOC wins and the distortions follow the SOC route. Such is, for example, the situation in FeO and CoO, or in KFeF$_3$ and KCoF$_3$. CoO is especially interesting, since there is in it a very large magnetostriction exactly due to this effect. Formation of the long-range magnetic order below $T_N \sim 300$~K results in cooperative lattice distortions with $c/a<1$ and a small thermal hysteresis at $T_N$\cite{Kanamori1957}.

Note that above we only considered tetragonal distortions. The $t_{2g}$ levels, however, can also be split by trigonal distortions. Experimentally most compounds of Co$^{2+}$ distort tetragonally,  and those of Fe$^{2+}$ - trigonally. Why is that, is not actually clear. 

\subsubsection*{Strong SOC: $d^4$ and $d^5$ configurations\label{JTSOC:45}}

 In typical $4d$ and $5d$ systems with their low-spin states, as well as for $3d$ systems with small number of $d$-electrons, where these electrons are only in the $t_{2g}$ levels, the situation is different. In this case one should not consider only terms like $\lambda l^z s^z$, but has to take into account the SOC ``in full force'' (with terms like $l^{+}s^{-}$ and $l^{-}s^{+}$), including also quantum effects. Eventual interplay of the JT effect and the SOC then looks different, and the results actually strongly depend on a particular situation, i.e. on the orbital occupation. For the general case, where the strength of the JT coupling ($E_{JT}$) and the SOC constant $\lambda$ are comparable, one has to carry out a special detailed treatment. But in the limit of strong SOC one can get some results rather easily qualitatively.

The first case to consider is the already discussed in Sec.~\ref{sec:SOC-intro} situation of the low-spin $d^5$ configuration, like that in Ir$^{4+}$. In the absence of the SOC this would correspond to one hole in triply-degenerate $t_{2g}$ levels, and the usual JT effect would lead e.g. to the tetragonal elongation (level structure is shown in Fig.~\ref{fig:JT-elong}a), with this hole on the $xy$ orbital (or there can be trigonal distortions, with a hole residing on the $a_{1g}$ orbital). However, in case of very strong SOC ($jj$ coupling) the splitting will be very different: the ground state of such ion is a Kramers doublet $J=1/2$, with the wave functions 
\begin{eqnarray}
|J_{1/2}, J^z_{1/2} \rangle = \frac 1 {\sqrt 3}\left( |xy \uparrow \rangle +  |(i xz + yz) \downarrow \rangle  \right),  \nonumber  \\           
|J_{1/2}, J^z_{-1/2} \rangle  = - \frac 1 {\sqrt 3}\left( |xy \downarrow \rangle + |(i xz - yz) \uparrow \rangle     \right),  \label{eq:j12}
\end{eqnarray}
see Fig.~\ref{jj-d4-d5}. Kramers doublets have no extra (orbital) degeneracy, i.e. there would be no JT effect in such state. Thus, we see that in this case the strong SOC completely suppresses the JT distortions (and vice versa, if we would make such distortions, e.g., elongation of MO$_6$ octahedron, then a hole would occupy the $xy$ orbital, which is the state with $l^z_{eff}=0$, i.e. such distortion would quench the SOC). It is interesting that while there is no orbital degeneracy in the ground state in case of large SOC for the $d^5$ configuration, it still exists in the excited state\cite{Plotnikova2016}. 

The situation for the $t_{2g}^4$ configuration in the case of a strong SOC is very similar. Without the SOC we would again have the orbital degeneracy, and corresponding JT distortion would be a tetragonal compression of MO$_6$ octahedra (or similar trigonal distortion), for which the $xy$ singlet  (or $a_{1g}$ for trigonal distortion) would go down and would be filled by two electrons, with the $xz, yz$ doublet  (or $e_g^{\pi}$ doublet) lying above with two electrons with parallel spins, Fig.~\ref{fig:JT-elong}b. But the SOC would prefer a very different orbital filling - the one shown in Fig.~\ref{jj-d4-d5} with the singlet ground state $J=0$. Thus, in this case the strong SOC will also suppress the JT effect.

\subsubsection*{Strong SOC: $d^1$ and $d^2$ configurations\label{JTSOC:12}}
However, the situation is different for $d^1$ and $d^2$ configurations. In this case, for less-than-half-filled $t_{2g}$ subshell, the third Hund's rule tells us that the order of multiplets is inverted, and the lowest would be a quartet $J=3/2$ (the same conclusion is also valid in $jj$ coupling scheme, see Sec.~\ref{sec:SOC-intro}). The one-electron states of $J=3/2$ quartet are two Kramers doublets:
\begin{eqnarray}
\label{3/2-states}
&&|J_{3/2}, J^z_{3/2}\rangle = |l^z=1, \uparrow \rangle = - \frac 1{\sqrt 2} (|yz, \uparrow \rangle + i |xz, \uparrow \rangle ),\nonumber \\
&&|J_{3/2}, J^z_{-3/2}\rangle = | l^z=-1, \downarrow \rangle 
=  \frac 1{\sqrt 2} (|yz, \downarrow \rangle - i |xz, \downarrow \rangle  ),
\nonumber \\
&&|J_{3/2}, J^z_{1/2}\rangle = \sqrt{\frac 2 3}|l^z=0, \uparrow \rangle - \frac 1 {\sqrt{3}}|l^z=1, \downarrow \rangle,  \nonumber \\ 
&&\quad \quad  =  \sqrt{\frac 2 3}|xy, \uparrow\rangle - \frac 1 {\sqrt{3}}|\frac 1{\sqrt 2} (yz + ixz), \downarrow \rangle, \nonumber \\
&&|J_{3/2}, J^z_{-1/2}\rangle = \sqrt{\frac 23}|l^z=0, \downarrow \rangle + \frac 1 {\sqrt{3}}|l^z=-1, \uparrow \rangle  \nonumber \\ 
&&\quad \quad =  \sqrt{\frac 2 3}|xy, \downarrow \rangle + \frac 1 {\sqrt{3}}|\frac 1 {\sqrt 2} (yz - ixz), \uparrow \rangle. \label{eq:j32}
\end{eqnarray}
In effect there is not only a Kramers degeneracy, but an extra (orbital) degeneracy  (note that it is not a triple degeneracy as in the original $t_{2g}$ shell, but  double degeneracy, two Kramers doublets!).  And this extra degeneracy can be again lifted by distortions, by the same JT effect.

Without SOC of course these configurations, $d^1$ and $d^2$, are JT active, leading to opposite distortions (tetragonal compression for $d^1$ and tetragonal elongation for $d^2$) Interestingly enough, one can show that in the case of strong SOC, for $\lambda \to \infty$ the still present JT distortion is should be such that both tetragonal elongation and compression give the same energy of the distorted state in the first approximation (this reminds formation of the ``Mexican hat’’ in the JT effect for doubly degenerate $e_g$; for the triply degenerate $t_{2g}$ situation is very different). The large but finite $\lambda$ would lead to the lowering of the energy of tetragonally compressed structure for the $d^1$ configuration, and to elongation for the $d^2$ configuration -- the same as for a pure JT effect for $\lambda=0$. However nonlinear effects, such as local anharmonism \cite{KhomskiiBrink2000} could change the situation. Thus we see that in this case even very strong SOC does not completely suppresses JT distortion, but still reduces it: due to the presence of Klebsch-Gordon coefficients $\sqrt{2/3}$ etc in the expressions \eqref{eq:j32}
wave functions for $J=3/2$ quartet) the JT energy gain turns out to be half of what one would get without the SOC.

\subsubsection*{Strong SOC: $d^3$ configuration}
Very unusual situation could exist for $d^3$ configuration,  for nominally half-filled $t_{2g}$ shell. In the usual $LS$ coupling scheme we would then have $S=3/2$ state with a quenched orbital moment $L=0$ and without the SOC (in the first approximation). However the situation would be very different for the strong SOC, in the $jj$ coupling scheme. It has been mentioned in Sec.~\ref{sec:SOC-intro} that for $d^3$ configuration these two coupling schemes, $LS$ and $jj$, lead in general to different states: the pure spin $S=3/2, L=0$  quartet in the $LS$ coupling scheme, and the spin-orbit-determined $J=3/2$ quartet in the $jj$ scheme, with different wave functions and generally speaking with different physical parameters such as the $g$-factors etc. But also from the point of view of the JT effect these two states are different. There is no orbital degeneracy left for the half-filled $t_{2g}$ shell in the $LS$ coupling scheme. However, it is not the case in the $jj$ scheme. Again, we have here {\it two} Kramers doublets $|J^z = \pm 3/2 \rangle$ and $|J^z = \pm 1/2 \rangle$, with different wave functions, \eqref{eq:j32}, and with different (opposite) local JT distortions. If we put three electrons on these states, one of these doublets would be necessarily filled, but another half-filled, so that the total distortions would not cancel, and such $d^3$ configuration would again be JT active and would lead to the JT distortion! We see that in this case, in contrast to the situation of Sec.~\ref{JTSOC:45}, the SOC does not suppresses, but activates, causes  JT distortion!  It would be very interesting to confirm these considerations experimentally. The absence/presence of the JT effect for the $d^3$ configuration could be a fingerprint of the applicability of the $LS$  (Russel-Saunders) or $jj$ coupling schemes for a particular material.

\subsection {Spin-orbit coupling and the formation of ``molecules'' in solids}

Similarly to the JT effect discussed in the previous section, the SOC can influence the formation of the MO states in solids, discussed in Secs.~\ref{sec:molecules} and~\ref{sec:OS}. And again, the detailed results depend on a particular situation.

Generally one should expect that the strong SOC would act against the formation of bonding states for example in TM dimers. Consider for example the case of the common edge geometry shown in Fig.~\ref{fig:packing}(b), for which  only the $xy$ orbitals can form bonding MO state (we ignore here possible hoppings between the $xz$ and $yz$ orbitals via oxygens). We gain maximum bonding energy when electrons occupy these $xy$ orbitals. But the SOC may favor very different orbitals.  For example in the case of $d^1$ configuration strong SOC would stabilize an electron on a $J=3/2$ quartet, Eq.~\eqref{eq:j32}. In this case the Kramers doublet $|J_{3/2}, J^z_{\pm 3/2}\rangle$  does not form bonding states at all  ($xy$ orbitals do not enter the states of this doublet). Only the doublet $|J_{3/2}, J^z_{\pm 1/2} \rangle$, containing the $xy$ component, would contribute to the bonding. Both these $xy$ orbitals enter to the $|J_{3/2}, J^z_{\pm 1/2} \rangle$ wave function with the coefficient $\sqrt{2/3}$, see \eqref{eq:j32}. Correspondingly, the bonding energy in this case would be reduced, it would be $-2/3t$ instead of $-t$ for the real  $xy$ orbital. Thus, we see that in this case the strong SOC leading to the formation of $J=3/2$ quartet, partially suppress the tendency towards the MO formation.

The same arguments would work not only for one electron, but also for one hole in the $t_{2g}$ shell, such as Ir$^{4+}$ ions. According to \eqref{eq:j12} in this case the ``active'' $xy$ orbital enters $J=1/2$ wave function with even smaller coefficient, $1/\sqrt 3$, so that in effect the bonding energy would be reduced even stronger, by factor 3: $E_{bond} = -1/3t$.

For other electronic configurations, such as, e.g., $d^2$ or $d^3$, the situation could be even more tricky. One has then also to worry about the role of the Hund's coupling. We have seen in Sec.~\ref{sec:OS} that the Hund's coupling counteracts the kinetic energy (hopping), i.e. it acts against MO formation. But in a general case it may also work against the SOC, since the Hund's exchange maximizes the spin $S$, while the SOC takes care about the total moment $J$. We will not discuss these different cases here; it is sufficient to say that the features of the MO formation in correlated materials, especially for those of the $4d$ and $5d$ elements, can be also sensitive to the SOC; and vice versa, strong intersite effects could in principle suppress the SOC.

\subsection{$J_{eff}=1/2$ and the spin-orbit assisted Mott state}
Usually when we go in the periodic table down in a column, e.g. from Co to Rh and to Ir, larger spatial extension of the $4d$ and $5d$ orbitals (compared with $3d$) and stronger covalency lead to a more metallic behavior. However, in the systems Sr$_2$MO$_4$ the tendency is the other way round. While Sr$_2$CoO$_4$ (difficult to prepare, but still), and Sr$_2$RhO$_4$ are metallic\cite{Matsuno2004,Kim2006}, Sr$_2$IrO$_4$ is insulating\cite{Crawford1994}. It turns out that one needs to take into account the SOC to describe this behaviour.

The electronic configuration of these TM ion is $d^5$. Due to alarge $t_{2g}-e_g$ crystal-field splitting all these electrons occupy $t_{2g}$ levels. As it has been shown in Sec.~\ref{sec:SOC-intro}, strong SOC lifts orbital degeneracy, and the ground state will be the Kramers doublet $J_{eff}=1/2$.  The situation will then become essentially equivalent to that of a half-filled nondegenerate Hubbard model, described by Eq.~\eqref{Hubbard-model}\cite{Kim2008}. The critical $U_c$ for the Mott transition for these $J_{eff}=1/2$ states is smaller than that for whole $d$-band, since, first of all, $U_c \sim \sqrt N$, \cite{Gunnarsson1996} where $N$ is the orbital degeneracy ($N_{J_{eff}}=1$, while $N_{t_{2g}}=3$). Second, the width of $J_{eff}=1/2$ band is smaller than the width of the whole $t_{2g}$ band\cite{Kim2008}. This explains why Sr$_2$IrO$_4$ is a (Mott) insulator, whereas Sr$_2$CoO$_4$  and Sr$_2$RhO$_4$ are metallic \cite{Kim2008}. This paper  started the whole activity in studying correlated solids with strong SOC, which has lead to some quite interesting and nontrivial results.

\subsection{Spin-orbit driven Peierls transition}
After discovery of strong influence of the SOC on the Mott transition it became clear that the SOC can be also very important for many other physical effects, e.g. for the Peierls effect. Indeed, these are the Peierls distortions which allowed to explain the highly unusual and seemingly self-contradictory properties of CsW$_2$O$_6$. In this compound W is 5.5+ and nominally has 1/2 electron per site. In spite of noninteger occupancy it is a nonmagnetic insulator in the low temperature phase (below 210K)\cite{Hirai2013}. Because of the large  $t_{2g}$ bandwidth the Hubbard correlations are helpless to explain this fact\cite{Hirai2013}. The solution of this problem came with the account of the SOC coupling, which strongly modifies the band structure making it susceptible to the Peierls transition\cite{Streltsov2016c}. The Fermi surface exhibits a strong nesting in this case, electronic susceptibility (Lindhard's) shows clear divergence at the same $\mathbf q$ vector. The calculations of the phonon spectra and the subsequent lattice optimization allowed us to find the crystal structure with tetramerized W-W chains running in two orthogonal directions in two different $ac$ planes. CsW$_2$O$_6$ turned out to be nonmagnetic band insulator in this picture, which fully explains all experimental observations.

While it is hard to see any 1D bands in CsW$_2$O$_6$ even taking into account the SOC, one might propose a very simple model, which explains the Peierls instability in this compound and the importance of the SOC. The $\beta-$pyrochlore structure of CsW$_2$O$_6$ strongly reminds spinel (with TM ions in the $B-$sites), which are prone to the Peierls distortions due to ``1D-zation'' of electronic spectrum, as we have seen in Sec.~\ref{sec:orb-Peierls}. Here WO$_6$ octahedra are elongated and thus electrons occupy two degenerate $xz/yz$ bands. The SOC lifts this degeneracy, and we have 1/2 electron in doubly degenerate (taking into account spin) band, which naturally explains tetramerization found in the band structure calculations.

\begin{figure}[t]
   \centering
  \includegraphics[width=0.32\textwidth]{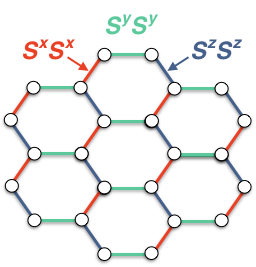}
  \caption{\label{fig:Honeycomb} Kitaev model on a honeycomb lattice.}
\end{figure}

\subsection{Kitaev exchange\label{sec:Kitaev}}
One of the most interesting consequences of strong SOC is the conclusion reached by Jackeli and Khaliullin\cite{Jackeli2009} that the $d^5$ systems with honeycomb lattice such as Li$_2$IrO$_3$, Na$_2$IrO$_3$ or $\alpha-$RuCl$_3$ might have very unusual type of exchange interaction, which is nowadays called Kitaev interaction. Instead of the Heisenberg model \eqref{eq:Heisenberg} it can be described by the Hamiltonian
\begin{eqnarray}
\label{eq:Kitaev}
H = \sum_{ij} K_{ij} S_i^{\gamma} S_j^{\gamma}.
\end{eqnarray}
For each bond this interaction has an Ising character, but with different $S$ components (numbered by the index $\gamma = \{x,y,z\}$) ``working'' on different bonds, see Fig.~\ref{fig:Honeycomb}. Such model, called there the compass model, was first introduced in \cite{KK-UFN} in treating orbital ordering, and anisotropic exchange there was caused by the directional character of orbitals, mentioned many times above in this review. Kitaev independently formulated this model in \cite{Kitaev2006}, and, most importantly, has shown that on a honeycomb lattice this model can be solved exactly, and the solution displays quite nontrivial states such as the spin—liquid state with short-range correlations, Majorana fermions etc. These results attracted enormous attention, see e.g.  \cite{Trebst2017}, especially because one could think of using the special properties of such systems for quantum computation\cite{Kitaev2006,Kitaev-new}. It was shown in \cite{Jackeli2009} that honeycomb materials with $t_{2g}^{5}$ electronic configuration (Ir$^{4+}$ or Ru$^{3+}$ ions) could be real examples of Kitaev systems.

\begin{figure}[t]
   \centering
  \includegraphics[width=0.49\textwidth]{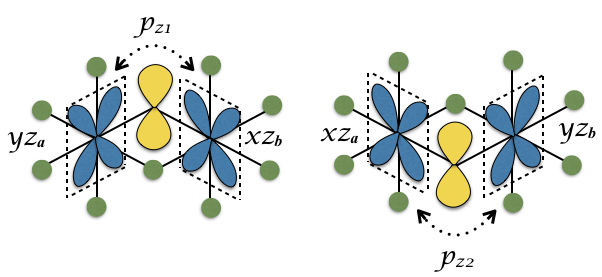}
  \caption{\label{fig:Kitaev} Possible exchanges paths between two $t_{2g}$ orbitals via ligand $p_z$ orbital in the common edge geometry.}
\end{figure}

The origin of the bond-dependent interaction \eqref{eq:Kitaev} is explained in Fig.~\ref{fig:Kitaev}. We have seen that a single hole resides on $J=1/2$ levels in case of strong SOC (Fig.~\ref{jj-d4-d5}). As one can see from Fig.~\ref{fig:Kitaev}, there are two equivalent passes for virtual hopping (via $p$ orbitals of ligands) from one Ir to another in edge sharing geometry, which usually leads to (antiferromagnetic) superexchange. However, for strong SOC, with the wave functions \eqref{eq:j12}, the total effective hopping between these wave functions \eqref{eq:j12}, $t_{dd}^{eff}$, exactly cancels due to the presence of $i$ for one of the relevant $d$-orbitals. This is why conventional superexchange given by~\eqref{t2U:SE}  does not work in this situation. If one assumes that the direct exchange between $t_{2g}$ orbitals is zero, then what remains is the higher-order processes (hopping to empty orbitals, with the Hund's rule acting there). These higher-order processes lead to the Ising-like interaction 
$K S_i^{z} S_j^{z}$   for the $xy$ plaquettes of Fig.~\ref{fig:Kitaev}, and to similar interactions with $K S^xS^x$ for the $yz$ plaquettes (bonds) and $K S^yS^y$ for the $zx$ bonds,  where $S$ is the effective spin $S=1/2$ for $J=1/2$ Kramers doublet \cite{Jackeli2009}. The exchange constant is of order 
\begin{eqnarray}
K \sim - \frac {t_{pd}^4}{\Delta_{CT}^2 U} \frac {J_H}U,
\end{eqnarray}
very similar to what we obtained in Eq.~\eqref{SE:hfe}.

For real honeycomb materials like Na$_2$IrO$_3$  or $\alpha-$RuCl$_3$ different Ir-Ir or Ru-Ru bonds have different orientation, see Fig.~\ref{fig:Kitaev}, so that in effect on three bonds going from each Ir or Ru ion we get the interactions of the type $S^zS^z$ for one bond,  $S^xS^x$ for the other, $S^yS^y$ for the third, i.e. the Kitaev model \eqref{eq:Kitaev}. 

The deviations from the exact cubic symmetry or from 90$^{\circ}$ metal-ligand-metal angle, as well as some other exchange processes, e.g. due to direct $d-d$ hopping of the $xy$ orbitals, would add to this interaction also some Heisenberg terms, so that the resulting model is of Heisenberg-Kitaev type. Also the exchange processes important for charge-transfer insulators, see \eqref{t2U:CT} and Fig.~\ref{fig:SU-t2}(b)  (with virtual states with two holes on one oxygen) would give Heisenberg terms in the exchange (each exchange pass, via each oxygen, acts here independently, so that there would be no interference terms and no cancellation of hoppings leading to the Heisenberg interaction; see also \cite{Jackeli2009}).

The question of the relative importance of Heisenberg or Kitaev terms, as well as the possible role (and the form) of more distant interactions for different real materials is a matter of active experimental and theoretical study, see e.g. \cite{Trebst2017}.

\subsection{Singlet (or excitonic) magnetism}
The situation with ions with the $d^4$ configuration for $4d$ and $5d$ materials with strong SOC  deserves special consideration. To such ions belong  for example Ir$^{5+}$ or Ru$^{4+}$. According to the treatment presented above such isolated ions in cubic crystal field should have a nonmagnetic singlet ground state with $J=0$. And indeed, Ir$^{5+}$ is the famous nonmagnetic ion: for ESR (electron spin-resonance) people it is a classical ion for nonmagnetic dilution. However, it is in principle still possible that there may exist magnetic state of such ions and even long-range magnetic ordering; for example it is typical for insulating Ru$^{4+}$ compounds, e.g. Ca$_2$RuO$_4$ and Na$_2$RuO$_3$ are antiferromagnets at low temperature\cite{Braden1998,Wang2014}. This may be a typical case  of a singlet magnetism, see e.g. Sec. 5.5 in \cite{khomskii2014transition}. Indeed, first of all the SOC may be partially quenched by lattice distortions, which lead to a noncubic crystal-field. Then, the exchange interaction with neighboring ions could be strong enough so as to overcome the initial splitting of the ground state nonmagnetic singlet $J=0$ and the excited triplet $J=1$: if Zeeman splitting of such triplet (by the internal exchange field from all other ions) exceeds the splitting between $J=1$ and $J=0$ state (given by $\lambda$), a magnetic state would have lower energy. This is the typical situation of singlet magnetism, well known for many rare earths compounds, e.g. those with Pr. Recently this topic became popular after asuggestion of G. Khaliullin \cite{Khaliullin2013} that many $d^4$ systems, for example, those with Ru$^{4+}$, can be described by this model; he called the resulting magnetic state an excitonic magnet. The phenomenological description of  the resulting magnetic state of materials like Ca$_2$RuO$_4$ is still possible with the usual exchange Hamiltonian for $S=1$, but containing strong anisotropy terms \cite{Kunkemoller2015,Komleva2017,Jain2017}, but there are also interesting new predictions such as for example the existence of a new spin-wave mode for ``soft'' spins, which may be called the Higgs mode. It seems to be observed in Ca$_2$RuO$_4$ in \cite{Jain2017}.

\section{Conclusions}
An interplay between the spin, charge and lattice degrees of freedom in transition metal compounds gives rise to various important physical effects such as giant magnetoresistance, high-temperature superconductivity and many others. An account of directional character of orbitals additionally enriches physical phenomena met in these systems. It turns out that in many cases orbitals play a role of either a transmitter, which establish a link between magnetic, electronic and elastic properties, or a tuner, which regulates interplay between them. As an example of the first ``emploi'' one may recall the Jahn-Teller effect, which couples electronic and elastic properties, or the presence of a magnetic anisotropy, which is usually related to the spin-orbit interaction. The second role of orbitals as a fine tuner becomes more and more important in last years. Thus, as we have seen, these are orbital degrees of freedom  that tune the exchange interaction in honeycomb systems like Na$_2$IrO$_3$ or $\alpha-$RuCl$_3$ and may result in Kitaev physics with spin-liquid ground state and highly unusual excitation spectra.  It is indeed rather interesting that Kitaev first solved his exotic model and found some nontrivial implications,  and only later on it was 
realized that the orbital degrees of freedom can tune a system to the regime where it can be described by this model. We expect that this second role of orbitals will become increasingly important, both since it opens new perspectives for the fundamental science and due to possible technological applications. In particular one might think of ``orbital engineering’’ on surfaces, interfaces etc.

Another tendency in the orbital physics, which has to be mentioned, is the change of a general route. Previously, most of the activity in this field was concentrated on the study of spin-orbital entanglement in Kugel-Khomskii-like Hamiltonians due to superexchange, on the analysis of the magnetic properties to different materials due to celebrated Goodenough-Kanamori-Anderson rule, or was connected with interplay between orbital and lattice degrees of freedom via the Jahn-Teller effect. In recent years a very different class of phenomena came to the forefront. First of all, these are the specific phenomena due to a directional character of orbitals. Second, this is the influence of orbital degrees of freedom on ``classical’’  effects, such as, e.g., Mott and Peierls transitions. Third, a lot of studies are now concentrated on the phenomena related to the spin-orbit coupling. We have found that the spin-orbit coupling can be important almost for all effects we know in the condensed matter physics: superconductivity, Jahn-Teller and Peierls effects, Mott transition, it results in a very similar spin-orbit entanglement as in the case of the superexchange and leads to pronounced exchange anisotropy. One may expect that the list of these phenomena will only widen in coming years.

In this review we tried to describe the novel development in the field of orbital physics. We hope that we demonstrated that this part of condensed matter physics, though not new, is still a very active field of research and is able to produce new and new surprises.

\vspace{5mm} 
\begin{center}* * *\end{center}
\vspace{5mm} 

Only a year ago we published a paper in the special issue of JETP devoted to the 85th birthday of Leonid Veniaminovich Keldysh. And unfortunately now we have to write a paper for the memorial issue of UFN. One of us, D. Kh., was one of his first PhD students, and later on for many year he was a member of his sector at the Department of Theoretical Physics at the Lebedev Physical Institute of the Academy of Sciences. And the interaction with L.V. Keldysh over many years was really crucial for his development.  Both of us express deep sorrow of the loss of L.V., and we are sure that the memory of L.V. Keldysh, both as a brilliant physicist and a wonderful person, would remain with us for many years to come.

This work was supported by the Russian Science Foundation through the project 17-12-01207.

\bibliography{../library}
\end{document}